\documentclass[preprint]{aastex}
\usepackage{emulateapj5}

\newcommand{\ksxrb}{\mbox{KS 1731$-$260}}
\newcommand{\slowb}{\mbox{4U 1728$-$34}}
\newcommand{\aqlx}{\mbox{Aql X-1}}
\newcommand{\scox}{\mbox{Sco X-1}}
\newcommand{\mxbgc}{\mbox{MXB 1743$-$29}}

\newcommand{\uxrbs}[2]{\mbox{4U #1$-$#2}}
\newcommand{\smstrut}{\rule[-1.0em]{0em}{1.5em}}
\newcommand{\twid}[1]{\mbox{$\sim$#1}}
\newcommand{\dddnuz}{\mbox{$\ddot{\nu}\dot{^{}}_0$}}
\newcommand{\etal}{\mbox{et al.}}

\newcommand{\nub}{\mbox{$\nu_{\rm b}$}}
\newcommand{\nud}{\mbox{$\nu_{\rm d}$}}

\newcommand{\nua}{\mbox{$\nu_{\rm a}$}}
\newcommand{\nus}{\mbox{$\nu_{\rm s}$}}

\newcommand{\taunu}{\mbox{$\tau_\nu$}}
\newcommand{\phiz}{\mbox{$\phi_0$}}
\newcommand{\Rapp}{\mbox{$R_{\rm app}$}}
\newcommand{\Tapp}{\mbox{$T_{\rm app}$}}
\newcommand{\Teff}{\mbox{$T_{\rm eff}$}}
\newcommand{\rxte}{{\it RXTE}}

\newcommand{\ergcms}{erg cm$^{-2}$ s$^{-1}$}
\newcommand{\ergsec}{erg s$^{-1}$}
\newcommand{\kmsec}{km s$^{-1}$}
\newcommand{\ctsec}{c s$^{-1}$}
\newcommand{\Msun}{\mbox{$M_\odot$}}

\newcommand{\mdotedd}{$\dot{M}_{Edd}$}

\begin{document}

\shortauthors{Muno \etal}
\shorttitle{Bursts and Oscillations from KS 1731$-$260}

\submitted{To be published in ApJ v. 542}
\title{Nearly Coherent Oscillations in Type~I X-Ray Bursts from \ksxrb}
\author{Michael P. Muno, Derek W. Fox, Edward H. Morgan}
\affil{Center for Space Research, Massachusetts Institute of
       Technology, 77 Massachusetts Avenue, Cambridge, MA 02139}
\email{muno@mit.edu, derekfox@space.mit.edu, ehm@space.mit.edu}
\and
\author{Lars Bildsten}
\affil{Institute for Theoretical Physics and Department of Physics,
       Kohn Hall, University of California, Santa Barbara, CA 93106}
\email{bildsten@itp.ucsb.edu}

\begin{abstract}

We present an analysis of the full set of {\it Rossi X-Ray 
Timing Explorer} (\rxte) observations of
\ksxrb\ made between 1996~August and 1999~February, concentrating on
the nine type~I X-ray bursts that were observed.  We find that the
bursts divide naturally into two populations: ``fast bursts'' occur on
the Banana Branch when the accretion rate is high and exhibit short
decay times, high peak fluxes, and radius expansion
episodes. ``Slow bursts'' occur in the Island State at lower
accretion rates, have lower peak fluxes, higher fluences, longer
decay times, and show no evidence of radius expansion.  All five
of the fast bursts, and none of the four slow bursts, show coherent
oscillations near 524~Hz.  Thus the mechanism that produces the burst
pulsations may well be related to the helium-rich burning process
indicated by the other properties of the fast bursts.

We perform in-burst phase connection of the burst pulsations, which
allows us to unambiguously characterize their frequency evolution.
That evolution exhibits a variety of behaviors, including a sharp
spin-down during one burst.  The frequency evolution exhibited by two
bursts that occurred 2.6~years apart is consistent with nearly the
same exponential-relaxation model; in particular, the asymptotic
frequencies of the two bursts differ by only $0.13\pm0.09$~Hz.  The
evolution during another burst, not modeled as exponential, shows a
maximum frequency which is $0.6\pm0.2$~Hz greater than the smaller of
these asymptotic frequencies.  Applying our phase models, we find that
the pulsations are spectrally harder than the burst emission, with the
strength of the pulsations increasing monotonically with photon
energy.  Coherently summing all of the burst signals, we find upper
limits of \twid{5\%} relative to the power of the main pulse on 
any modulation at 1/2 or 3/2 the main pulse frequency.
We discuss our results in the context of models in which the burst
pulsations are produced by a hot spot on the surface of a spinning
neutron star.

\end{abstract}

\keywords{stars: individual: KS 1731$-$260 --- stars: neutron ---
stars: rotation --- X-rays: bursts --- X-rays: stars}

\section{Introduction}

\ksxrb\ is a transient X-ray binary system located near the galactic
center, discovered in August~1989 using the imaging spectrometer
aboard the {\it Mir-Kvant\/} observatory \citep{sun89}.  Three type~I
X-ray bursts were observed from the source during 24 1000~s
observations spanning 15~days \citep{sun90}, which established that
the source was a neutron star low mass X-ray binary (LMXB). Subsequent
ROSAT observations confirmed that \ksxrb\ was a typical neutron star
LMXB \citep{bar98}.  The source has been detected once in hard X-rays
(35--150 keV) by SIGMA, making \ksxrb\ one of several neutron star
LMXBs with states similar to the low-hard states of black hole
binaries, but during several other observations no hard tail was
detected \citep{bar92}. Recently, attention has focused on \ksxrb\
because it is one of only four sources displaying both coherent
oscillations during a type~I X-ray burst (at 524~Hz; \citet{smb97}) and
twin kilohertz quasi-periodic oscillations with an approximately constant
separation in Fourier frequency; in \ksxrb\ the measured
frequency difference is close to half of the burst frequency
\citep{wk97}.

In this paper we report on a comprehensive analysis of the nine X-ray
bursts that were observed during observations of \ksxrb\ with the
{\it Rossi X-ray Timing Explorer} (\rxte) between 1996~August and
1999~February.  The bursts occurred while the persistent luminosity
varied systematically, which allows us to study how the properties of
the bursts correlate with the spectral states of the source. Five of
the bursts exhibited coherent oscillations, and we systematically investigate
what conditions correspond to the presence of these
oscillations. Finally, we have tracked the evolution of the phases of
the burst oscillations, which allows us to place tight constraints on
any long-term change in oscillation frequency, to derive upper limits
on the strength of sub-harmonics and harmonics of the observed burst
frequency, and to determine the strength and energy spectrum of the
pulsations.

To place this work in context, we begin by reviewing briefly the
phenomenology and theory of type~I X-ray bursts. In Section~1.1, we
examine how the properties of bursts have been observed to correlate
with the properties of the persistent emission in several sources, and
how theories address these correlations.  In Section~1.2, we discuss
the observations of nearly coherent oscillations during type~I X-ray
bursts and their possible connection with the spin period of the
underlying neutron star. The remainder of the paper concerns the
analysis (Section~2) and interpretation (Section~3) of \rxte\
observations of \ksxrb.

\subsection{The diversity of type~I X-ray bursts}

Type~I X-ray bursts have been observed from more than 45 low-mass
X-ray binary systems (LMXBs; see \citep{lvt93} for a review, and \citep{coc98}
for some recent discoveries). These
bursts occur when unstable nuclear burning of helium or hydrogen
ignites matter accreted onto the surface of the neutron star primary.
The models reproduce the observed rise times (\twid{1}~s), durations
(\twid{1}~minute), and recurrence times (\twid{hours}) of bursts,
spectral softening in the burst tails, and total energies of the
bursts ($10^{39}-10^{40}$ ergs).  Detailed models predict variations
in the frequency and strength of bursts from a single source due to
changes in the composition of the burning material, which is in turn
determined by the metallicity of the matter accreted onto the neutron
star, the amount of hydrogen burned during the time between bursts,
and the amount of fuel left-over from the previous burst
\citep{fhm81, aj82, fl87, fuj87, bil97}.
In addition, variations from source to source are expected because of
variations in the core temperatures of the neutron stars and accretion
rate \citep{aj82, fl87}.	

The theoretical models that are most readily compared to observations
of burst sources are those that predict how burst properties in an
individual system should change as the accretion rate onto the neutron
star varies.  Of the forty-odd bursting LMXBs, only about ten have had
their burst and persistent emission properties correlated; for the
majority of sources, either few bursts are seen, or several bursts are
seen in a single luminosity state \citep{lvt93}.  Nonetheless,
studies of these ten sources reveal some intriguing relationships
between the persistent emission and the recurrence times, durations,
and energetics of the bursts which can be compared to theoretical
models.

The model of \citet{fhm81}
predicts that X-ray bursts should occur in three regimes depending on
the accretion rate ($\dot M$), 
which they label Cases~1, 2, and 3. We denote these accretion rates in
units of the Eddington accretion rate, \mdotedd. 
At the lowest
accretion rates ($< 0.01$\mdotedd, Case~3), the temperature in the
burning layer is too low for stable hydrogen burning; the hydrogen
ignites unstably, in turn triggering helium burning, which produces a
type~I X-ray burst in a hydrogen-rich environment. At higher accretion
rates ($>0.01$\mdotedd, Case~2), hydrogen burns stably via the hot CNO
cycle and is converted to helium as quickly as material is accreted
onto the neutron star. A pure helium layer develops at the base of the
accreted material, and heats steadily until a pure helium burst is
triggered. At higher accretion rates ($> 0.1$\mdotedd, Case~1),
material is accreted faster than it can be consumed by hydrogen
burning (which is limited by the rate of $\beta$-decays in the CNO
cycle), so that the helium ignites unstably in the H-rich environment
(see Bildsten 1998 for dependences of these accretion rates on the
metallicities). At higher accretion rates ($>$\mdotedd), stable helium
burning becomes important on the surface of the neutron star, which
depletes the primary fuel reserves and causes bursts to occur less
frequently. 

\citet{hf82} point out that the accretion rate
inferred from the luminosity of any 
given neutron star LMXB that exhibits bursts can
range from $\sim 0.01-0.1$ \mdotedd, so that an
individual source that varies in luminosity should exhibit changes in
its bursting behavior between Case~2 and Case~1. This model provides
three predictions.

First, as the accretion rate increases, the column of material above
the burning layer builds more quickly, and thus the time required to
reach the critical temperature for unstable helium burning
decreases. This is consistent with the increase in burst rate with
persistent flux observed in \uxrbs{1728}{34} \citep{bas84} and
\uxrbs{1820}{30} \citep{cla77}. At the highest accretion rates,
steady helium burning reduces the amount of fuel for X-ray bursts,
which causes bursts to occur less often, as seen in EXO~0748$-$676
\citep{got86} and \uxrbs{1705}{44} \citep{lan87}, or not at all, as
observed in \uxrbs{1820}{30} \citep{cla77} and GX~3$+$1
\citep{mak83}.  However, no correlation was found between persistent
flux and burst recurrence times in Ser~X-1 \citep{szt83},
\uxrbs{1735}{44} \citep{lew80, par88}, and \uxrbs{1636}{53}
\citep{lew87}, which suggests either that an additional mechanism may
control how frequently bursts occur, or that the persistent flux is not a good
measure of the accretion rate in these sources (see below).

Second, the ratio of the energy released in the bursts to that
released by accretion between bursts should be larger when large
amounts of hydrogen are burned during a burst, and should decrease as
more helium is added to the nuclear burning layer. Assuming that all
of the accreted material is eventually consumed in a burst, this ratio
is simply the amount of energy released by the nuclear reactions ---
the CNO cycle for hydrogen burning, and the triple-$\alpha$ process
for helium burning --- divided by the gravitational energy released by
the accreted material. Historically, the inverse of this ratio is
used, and defined as $\alpha$. The expected values for $\alpha$ are
$\sim25-100$, where the lower value corresponds to pure hydrogen
burning, which releases about four times the energy per nucleon than
helium burning does.  The value of $\alpha$ was observed to increase
as expected (from 12 to 75) as the persistent flux from EXO 0748$-$763
increased by a factor of five \citep{got86}. However, several sources
exhibit $\alpha$ values as large as $10^3$, which suggests that both
hydrogen and helium burn steadily during the periods of steady
accretion between bursts \citep{ppl88, bil95}.

Third, the larger the fraction of helium in a burst, the faster nuclear
energy will be released (via the triple-alpha process) during the helium
flash, so that the peak flux of the burst will be higher
\citep{fhm81, aj82}. Conversely, if more hydrogen is present
in the burning material, as in Cases~1 and 3, then helium will be
diluted, slowing the rise of the burst.  The hydrogen will, however,
burn slowly in the tail of the burst via proton capture onto the
products of the helium burning.  Thus the fastest, most intense bursts
should occur under Case~2, which corresponds to low accretion rates.
Surprisingly, this is precisely contrary to the sense of the majority of
the observations.  The decay time scales of bursts from
\uxrbs{1608}{52} \citep{mur80}, \uxrbs{1636}{53} \citep{lew87}, and
\uxrbs{1705}{44} \citep{lan87} have all been reported to decrease
with increasing flux.  The peak fluxes of bursts were found to
increase as the persistent flux increased in EXO~0748$-$676
\citep{got86}, \uxrbs{1608}{52} \citep{mur80}, and Ser~X-1
\citep{szt83}, although \uxrbs{1705}{44} \citep{lan87} and
\uxrbs{1735}{44} \citep{lew80, par88} exhibit no apparent correlation
between the peak fluxes of bursts and the persistent flux.
A survey of ten LMXBs by 
\citet{ppl88} reveals a global decrease in burst duration
with increasing flux. The sense of these observations is that the
fastest, most intense bursts occur at high accretion rates. The origin
of this discrepancy is uncertain (see Bildsten 2000 for a recent
discussion). 

It is apparent that for each correlation mentioned above, only a few
sources can be cited as examples. This is largely because two
systematic difficulties interfere with observation of these effects.
First, recurrence times and $\alpha$ values for bursts are difficult
to measure because most satellites observe from low Earth orbit, where
the Earth occults most sources for a significant fraction of every
hour.  Since bursts recur on a time scale of hours, many bursts are
missed during occultation. Only EXOSAT, with its 91~hour orbit, was
relatively free from this difficulty.

Second, the persistent flux from an LMXB is not a monotonic function
of accretion rate. As the trends predicted by the model of Fujimoto,
Hanawa, \& Miyaji (1981) are a function of the accretion rate, it is
not surprising that correlations between burst properties and
persistent flux would not be evident in every source. A better measure
of accretion rate is provided by the color-color diagram
\citep{vdk95}. Indeed, in the atoll source \uxrbs{1636}{53} both the
burst duration and the apparent temperature of the emission in the
burst tails were found to be correlated with the position on the
color-color diagram, but not with the flux
\citep{vdk90}. Observations using instruments prior to the EXOSAT
mission did not provide sufficient timing and/or spectral information 
to deduce the spectral state of LMXBs on a
color-color diagram unambiguously. Fortunately, the excellent time and
energy resolution available with \rxte\ allows us to examine the
relationship between burst properties and accretion rate in 
sources such as \ksxrb.

\subsection{Burst pulsations and kilohertz quasi-periodic oscillations}

Although type~I X-ray bursts are useful measures of the energetics, 
time scales and recurrence frequencies of nuclear burning on neutron stars, 
the bursts are only observed after the energy has
propagated through the neutron star envelope, and so they are not direct
probes of the burning layer itself.  The discovery of near-coherent
oscillations during type~I bursts from several sources using data
from the Proportional Counter Array (PCA; \citep{jah96} of the {\it
Rossi X-ray timing Explorer} ({\it RXTE}; \citet{brs93}) has opened a
new and extremely valuable window onto X-ray bursts and the neutron
stars underlying them (see \citep{ssz98} for a review).  The
oscillations have been seen in both rising and cooling phases of the
bursts and have extremely high coherence values ($\nu/\Delta\nu >
300$), although the pulse frequency often drifts by up to 2~Hz over
the course of the burst \citep{str98b, sm99}.  In particular,
the oscillations sometimes relax towards an asymptotic frequency which
is identical in bursts separated by several years --- stable enough
(to less than one part in one thousand) that it has been suggested it
might be used to constrain the binary mass function \citep{str98b}.

The obvious candidate for a pulsation of this high coherence and
long-term stability is the spin period of the neutron star
\citep{str96}.  Under this interpretation, it is usually proposed
that the modulation is the result of localized burst emission (a ``hot
spot'') on the neutron star surface.  Several lines of argument
support this hypothesis.  First, the fractional amplitudes of the
oscillations in the rise of bursts are generally strongest on the
leading edge of the bursts, and are undetectable at the burst peak
when a large portion of the neutron star surface is in conflagration
\citep{ssz98}. Second, the RMS strength of the oscillations in the
rise of the bursts is very high, up to $\sim$30\% in \slowb\
\citep{szs97} and $\sim$50\% in \uxrbs{1636}{53} \citep{str98c}; note
that in both cases the authors report the half-amplitude, which is a
factor of 1.4 larger than the RMS amplitude). This would be expected
if the initial burning is confined to a small area on the neutron star
surface \citep{ml98}.  Finally --- although it is not understood how
a hot spot could survive the initial conflagration on the neutron
star --- pulse phase-resolved spectroscopy of an oscillation seen in
the tail of a burst from \uxrbs{1636}{53} indicates that the flux
modulation is accompanied by a modulation of the blackbody temperature
of the spectrum, again as expected (Strohmayer \etal\ 1998a). A
similar conclusion has been reached for a burst from \aqlx\ through
the detection of soft lags in the burst photons, which are attributed
to Doppler shifts of the light from the fast-moving hot spot 
\citep{for99}.

The favored model for the drifts in frequency is that they indicate an
expansion of the burning layer by 10--30~m at the burst start. During
the expansion, the rotation of the burning layer slows, only to relax
back to the neutron star surface and spin up again later in the burst
\citep{str98b}. Cumming \& Bildsten (2000) have shown that the 
hydrostatic expansion expected during the burst is consistent with
that observed, and that the shear layers are likely stabilized by the 
strong stratification in the atmosphere. Consistent with this hydrostatic
model, \citet{str99} observed a decrease
in frequency of the burst pulsations during the tail of a burst from
\uxrbs{1636}{53} which was coincident with a re-heating episode.

Still, there are several unanswered questions regarding the hot spot
interpretation. Even in sources where the burst pulsations are seen,
not all bursts show them, which suggests that the creation of a hot
spot requires special conditions. (Until now, no systematic search for
correlations between the presence of coherent oscillations and the 
properties of the bursts or the persistent emission has been carried out.)
Moreover, as mentioned above, it is
not known how a hot spot could remain on the surface of the neutron
star after the initial burning phase, which presumably envelops the
whole neutron star \citep{szs97}.  Finally, \citet{mil00} finds that in one burst from \uxrbs{1636}{53}
(his Burst C) the pulsations have an ``asymptotic'' frequency that is
significantly less than the highest frequency observed during the
burst, which is hard to reconcile with the simple picture of a burning
layer contracting and re-coupling to the underlying neutron star.

The importance of the burst oscillations has been further enhanced by
the detection of quasi-periodic oscillations (QPOs) in the
200--1200~Hz frequency range in power density spectra (PDS) of the
persistent emission from numerous LMXBs.  These so-called kHz QPOs
(for reviews, see \citet{vdk97}, \citet{vdk00}) may reflect the
relativistic motions of material at the inner edge of the accretion
disk in these systems \citep{str96, mlp98, sv99}.  The
relative ubiquity of pairs of kHz QPOs --- with the twin peaks
separated by a frequency difference \nud\ that stays roughly constant
as the frequencies of the peaks increase \citep{str96, ketal97} 
--- leads naturally to beat-frequency models which
identify \nud\ as the spin frequency of the neutron star, \nus
\citep{str96, mlp98}.

Initially this interpretation was supported by the observation that
\nud\ was consistent (within errors) with the burst frequency \nub, or
half that, in sources where both phenomena were observed.  This
coincidence was evident in \slowb\ and \uxrbs{1702}{42}\
($\nud\approx\nub$), and in \uxrbs{1636}{53}\ and \ksxrb\
($\nud\approx\nub/2$).  (Note that in the burst-pulsation sources
\aqlx\ and \mxbgc\ twin kHz peaks have not been detected to date.)
The detection of a sub-harmonic signal in \uxrbs{1636}{53}\
\citep{mil99} has lent the idea further support by demonstrating that
the strongest signal in the bursts from this source is actually at the
first harmonic of the spin period, $\nub=2\nus$, so that
$\nud\approx\nus$.

However, it has also since been observed that (1) \nud\ is not
constant in \scox\ \citep{ketal97}, \uxrbs{1608}{52}\ \citep{men98},
and \slowb\ \citep{mk99}; (2) kHz QPO data from all other sources are
equally consistent with a changing or a constant \nud\ \citep{psa98};
and, (3) the ``equalities'' in \slowb\ and \uxrbs{1636}{53}\ do not in
fact hold ($\nud<\nub$ in \slowb, \citep{mk99}; and $\nud<\nub/2$ in
\uxrbs{1636}{53}, \citep{mkp98}.

One response to this dilemma is proposed by 
\citet{ot99}, who suggest that the decreasing difference
frequency is that of a differentially rotating magnetosphere. In their
model, the lower kHz QPO represents the Keplerian frequency at the
inner disk, and the upper kHz QPO occurs at an upper hybrid frequency
of Keplerian blobs in elliptical orbits, with the frequency determined
by the Coriolis force acting in the frame of reference rotating with
the neutron star magnetosphere.  Another set of models states that the
upper and lower kHz QPOs are the general relativistic Keplerian and
periastron precession frequencies near the inner edge of the accretion
disk \citep{sv99}.  The difference frequency in these latter models is
unrelated to the spin period of the neutron star, and the apparent
relationship between \nud\ and \nub\ is taken to be a coincidence.

It is therefore critical to determine the nature of the coherent
oscillations observed during bursts in order to distinguish between
models for the kHz QPOs. \ksxrb\ provides crucial observational
evidence for testing the current models, as it is one of only four
sources for which both the burst frequency and the difference
frequency have been measured.

\section{Observations and Data Analysis}

We have analyzed all 27 observations of \ksxrb\ taken between 1996
July~14 to 1999 February~27 with the Proportional Counter Array aboard
\rxte\ (Table~\ref{obs}).  During these observations, nine X-ray
bursts are seen, including the one previously reported by \citet{smb97}.

\placetable{obs}

\subsection{Characterization of persistent emission and source state}

\begin{figure}
\plotone{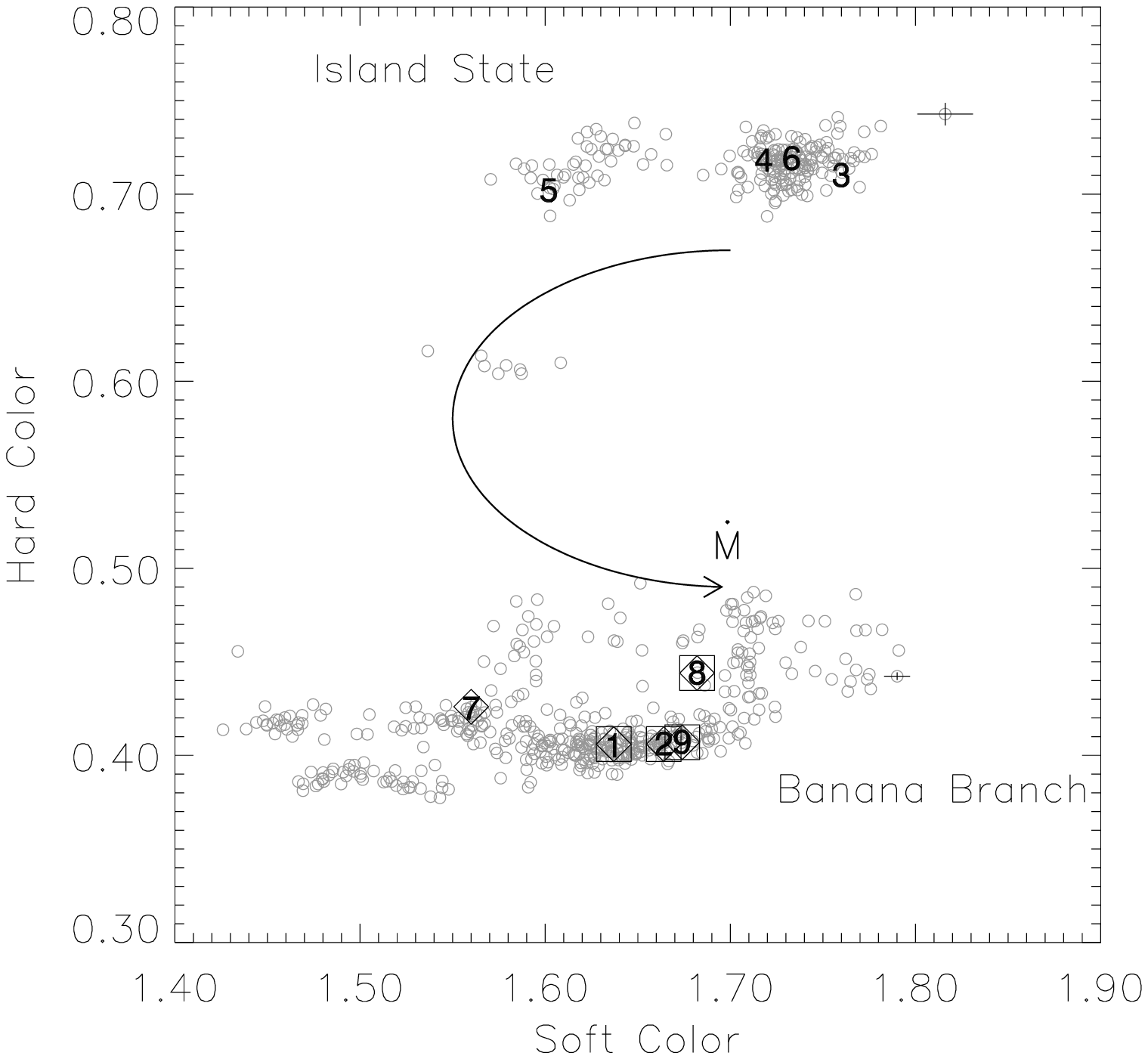}
\vspace{-1in}
\caption{Color-color diagram from PCA observations of KS 1731-260. The
Hard Color is the ratio of the count rates in the 8.5--18.0 keV and 
4.8--8.5 keV bands, while the soft color is the ratio of the count 
rates in the 3.4--4.8 keV and 2.0--3.4 keV bands. Background subtraction
was applied before calculating the ratios. Representative 1-$\sigma$
error bars are plotted at the right of the figure. The numbers
indicate the colors just prior to each of the nine bursts. Squares are placed
around bursts which exhibited radius expansion, while diamonds were placed 
around those which coherent oscillations. The arrow on the figure indicates 
the direction of increasing accretion rate.}
\label{cc}
\end{figure}

In order to characterize the persistent emission from \ksxrb, we have 
created background-subtracted light curves of the X-ray intensity of
\ksxrb\ from Standard2 data for each observation. We have calculated the
ratios of the count rates in the 3.4--4.8 keV $:$ 2.0--3.4 keV bands
(soft color), and in the 8.5--18.0 keV $:$ 4.8--8.5 keV bands (hard
color) for 256 s interval during each observations. The mean count
rates and hardness ratios, excluding time intervals containing bursts,
are listed in Table~\ref{obs}.  Figure~\ref{cc} displays the
color-color diagram for this set of observations.  The states with the
lowest observed count rates correspond to points with a hard color of
$> 0.65$, while those points with higher count rates have a hard color
$< 0.50$. The count rate from the source is not a monotonic function
of position on this diagram (Table~\ref{obs}). Although motion on the
branch with lower hard color is observed in the course of several
hours, the transition from a hard color $> 0.65$ to $< 0.50$ 
occurred between observations separated by more than a month.

\begin{figure}
\plotone{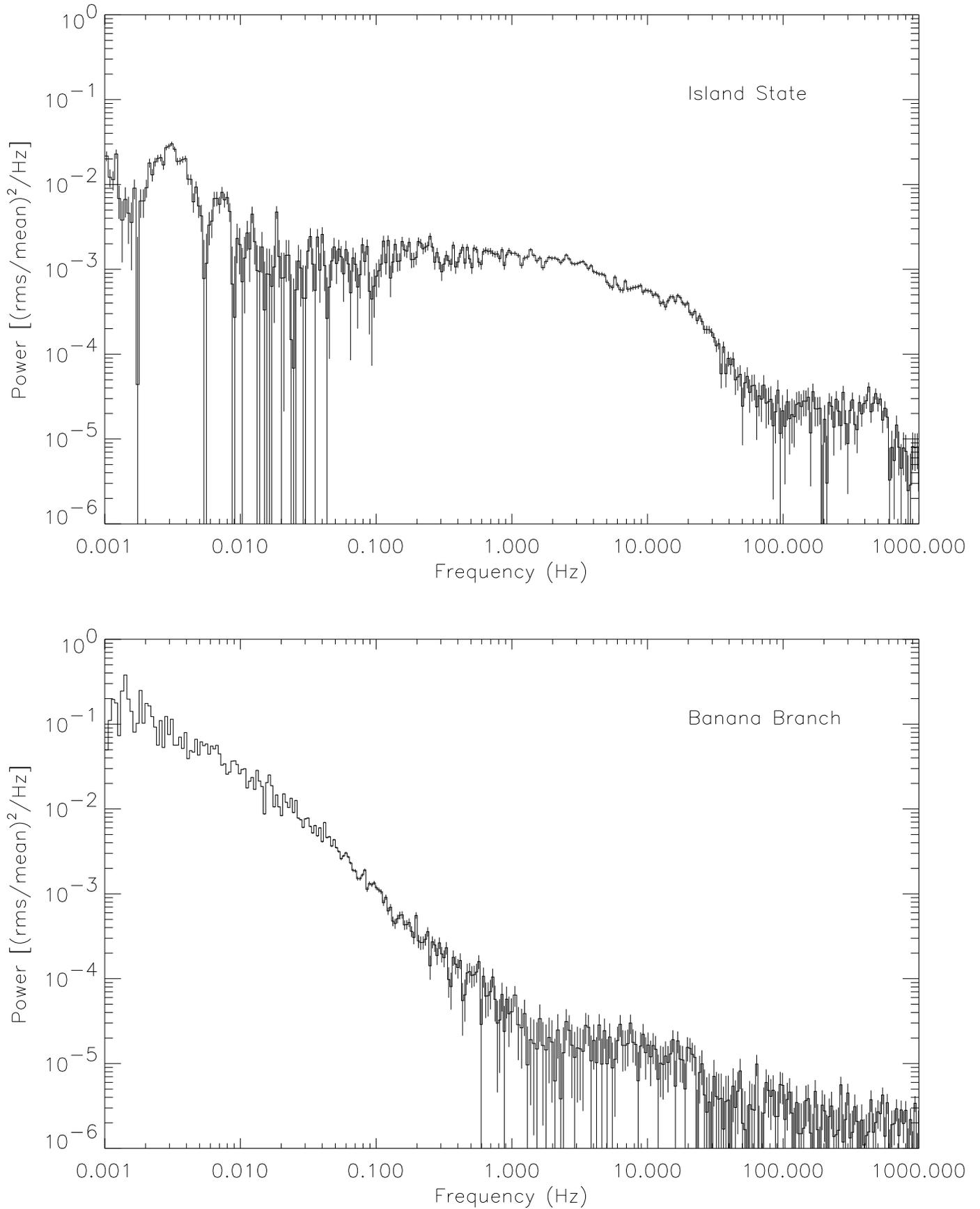}
\caption{Representative power density spectra from (a) the Island State 
(1998 October 6; 15000s), and (b) the Banana Branch (1997 October 28;
13000 s). The error bars represent the standard deviation in the powers
which were rebinned for each point in the above plot.}
\label{pds}
\end{figure}

\placefigure{cc}
\placefigure{pds}

We have produced power-density spectra (PDS) of photons in the 
2--60~keV \rxte\ band for each observation, in order to
determine whether \ksxrb\ exhibits the canonical branches of the
color-color diagram for LMXBs \citep{vdk95}.  The PDS of each
observation have been created by combining averaged 
PDS from data with $2^{-13}$~s time
resolution in 64 s intervals with a single 
PDS of data with $2^{-3}$~s time resolution for the entire observation. 
Bursts are excluded from the PDS.  We have found that PDS are of two 
types, as represented
in Figure~\ref{pds}.  Figure~\ref{pds}a is representative of PDS from
observations with hard colors $> 0.6$, which can be described as the
sum of relatively weak (2\% RMS) low frequency noise below 0.1 Hz and 
strong (13\% RMS) flat-topped high frequency noise that decreases in power 
above a few Hz. Figure~\ref{pds}b is representative of PDS from observations 
with
hard colors $< 0.5$, and which can be characterized as a combination
of $\sim 8$\% RMS low frequency noise below 1~Hz and weak power 
(2\% RMS) at high frequencies.
Strong ($> 1$\% RMS) QPOs are not evident below 100 Hz in the PDS from 
either state, and although single kHz QPOs are evident on a few 
occasions in photons with energy above 5 keV (not shown), twin kHz 
QPOs are only evident in the observation reported on by Wijnands \& Van 
der Klis (1997).

Taken together, the color-color diagram and PDS indicate that \ksxrb\
is indeed an atoll source, as suggested by 
\citet{wk97} (for a review of Z and atoll LMXBs, see
\citep{vdk95}; see \citep{mk99} for a similar color-color diagram of the
atoll source \slowb\ as observed with the PCA on \rxte).  The state
with hard color $> 0.65$ is the Island State, which is thought to
represent a low accretion rate ($\sim$ 0.01 times the Eddington rate,
${\dot{M}}_{Edd}$), while the state with hard color $< 0.50$ is
the Banana State, thought to occur at a somewhat higher accretion rate
($\sim 0.1 {\dot{M}}_{Edd}$). This classification of \ksxrb\ is
consistent with the presence of strong X-ray bursts --- only two
Z-sources, GX 17$+$2 \citep{szt86} and Cyg X$-$2 \citep{kk95},
exhibit type~I X-ray bursts. An arrow has been drawn on the color-color
diagram to indicate the direction believed to correspond to increasing
accretion rate.

We have estimated the source flux during each observation
(Table~\ref{obs}) by fitting the Standard2 data summed over each
\rxte\ orbit with a two-component model, consisting of an
exponentially cut-off power-law ($N_{\Gamma} E^{-\Gamma}\exp{-E/kT}$,
where $E$ is the photon energy, $\Gamma$ is the power law photon
index, $kT$ is the cut-off energy, and $N_{\Gamma}$ is the photon flux
at 1~keV) and a Gaussian line at 6.4~keV, which is likely due to iron
emission. We have accounted for interstellar absorption with a
low-energy cut-off equivalent to a column density of
10$^{22}$~cm$^{-2}$ of hydrogen (according to the value derived with
{\it ROSAT}, \citep{ps95}. The time intervals during bursts have been
excluded from the spectra. Although we fit the spectra between 
2.5--25 keV, the flux was calculated from the model between 2--18 keV, to 
correspond with the energy range used for the color-color diagram. 
The reduced chi-squared value for this
model is less than 1.1 for all observations except
\mbox{20058-01-01-00}, \mbox{30061-01-03-00}, and
\mbox{30061-01-04-01}, for which the reduced chi-squared values are
near two. The derived flux is not significantly affected by the poor
values of reduced-chi squared, and we report all values in
Table~\ref{obs}. For a more detailed discussion of the X-ray spectrum
of \ksxrb\ as observed by \rxte\, see \citet{bar99}.

\subsection{Characterization of type I X-ray bursts}

\placetable{burst}

Table~\ref{burst} lists the times (corrected to the Earth-Sun
barycenter) of the nine bursts observed with the PCA. We have
integrated spectra for each 0.25~s interval during these bursts using
combinations of the binned and event mode data for each
observation. This provides 32 energy channels for the first burst and
74 energy channels for the remainder of the bursts. Since the effect
of the X-ray burst on the spectrum of the persistent emission is
unknown, it is not clear what ``background'' should be subtracted from
the total received flux in order to obtain the true burst
spectrum. Therefore we have estimated the background in two
independent ways, representing two extremes of approach: (1) we have
used the FTOOL pcabackest to determine the instrumental and cosmic
background, including both the persistent and burst flux in our fits;
and (2) we have used the spectrum of 100~s of data prior to the burst
as background, subtracting the persistent emission under the
assumption that it is unaffected by the burst.  (We note that it is
possible that persistent emission increases during the burst, with
accretion induced by the burst itself, but we have no theoretical
guidelines for modeling such a process and therefore ignore it.)  We
have fit the resulting background-subtracted spectra between 2.5--25 
keV with a single blackbody (without absorption, since fits with absorption
favor very small values of the column density and do not improve 
chi-squared significantly), which provides two parameters: the
apparent temperature \Tapp\ and solid angle of the emission
region. From the solid angle we can derive the apparent radius \Rapp\
of the emission region given a derived or assumed distance to the
source. From the model parameters, we calculate the bolometric flux 
for each interval during the burst, $F = \pi R_{app}^2 \sigma T_{app}^4$.

We have compared spectral fits after subtracting the background in
these two ways, and find that the resulting parameters are very
similar, although using the pcabackest estimate of the background
increases $\chi^2_{\nu}$ for the fit by a factor of 1.3--1.6 relative
to fits with the pre-burst emission subtracted. The increase in 
$\chi^2_{\nu}$ is almost certainly due to the presence of flux from
steady accretion in the spectra that remain after subtracting the 
pcabackest estimate. To keep the spectral model simple, {\it we
report results only for spectra obtained after subtracting the
pre-burst emission.}  The resulting spectral parameters should be
interpreted with some caution, because it is not likely that the burst
emission is a pure black body: on the contrary, electron scattering in
the neutron star atmosphere is likely to cause systematic deviations
in the spectrum which would result in systematically high \Tapp\ (and
thus low \Rapp) at a given flux (see also the discussion in
Section~3).

We have analyzed the profile and strength of the bursts by finding the
rise time $t_{rise}$, start time of the decay $t_s$, decay time scale
$t_d$, maximum flux $F_{peak}$, fluence $E_b$, and duration $\tau =
E_b/F_{peak}$ for each burst (Table~\ref{burst}). The rise time
$t_{rise}$ has been defined as the time for the burst to increase from
25\% to 90\% of the peak flux. The start time of the decay $t_s$ is the
approximate time when the flux begins decreasing exponentially,
measured from the start of the burst.  We have fit the decay in flux
of each of these bursts with either one or two successive exponential
functions, $\exp(-t/t_{d})$, (for four and five bursts, respectively, 
depending on the value of reduced chi-squared for a fit with a single 
exponential) in order to estimate the time-scale for the decay of the 
bursts.  The
exponential fits work well; for all of the bursts, $\chi^2_{\nu} <
1.3$. Finally, we have summed the flux from the burst up to the start
of the burst decay with the flux in the burst tail, as determined by
integrating the exponential fit, to determine the fluence of the burst
$E_b$.

\begin{figure}
\plotone{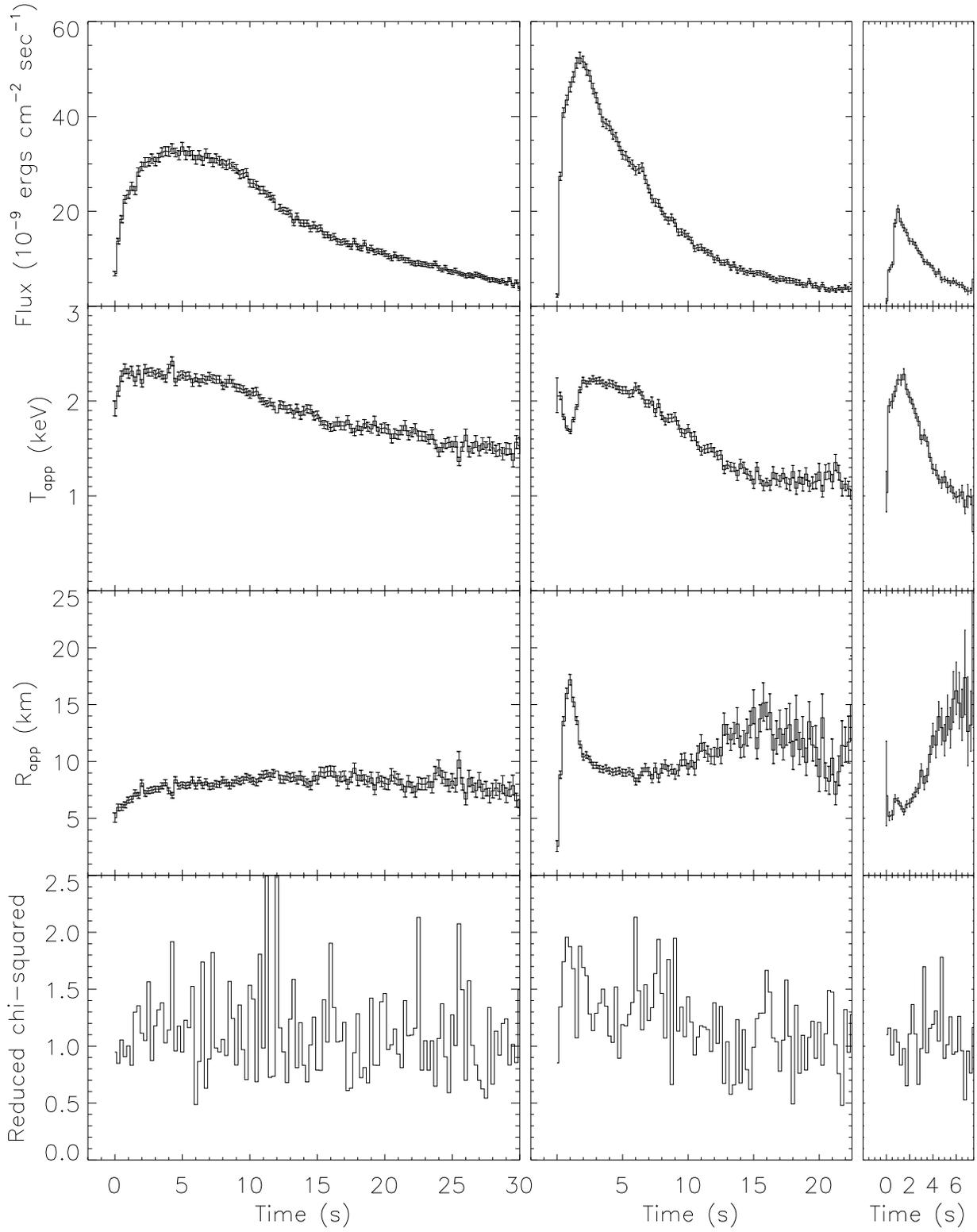}
\caption{Representative spectral fits to three bursts: (a) Burst 3, 
a long burst with no radius expansion; (b) Burst 9, with radius expansion;
(c) Burst 7, the only short, weak burst. From top
to bottom are bolometric flux, apparent temperature, apparent radius (assuming
a distance of 7 kpc; see text), 
and reduced chi-squared for the fit. Error bars plotted are 1~$\sigma$.}
\label{bspec}
\end{figure}

\placefigure{bspec}

The spectral fits reveal which of the bursts exhibited an episode of
radius expansion (last column in Table~\ref{burst}). We have found
that four bursts exhibit radius expansion episodes. In all cases the
effective area of the emitting region expands by a factor of
2.5--5.2. Since the maximum luminosity during the radius expansion
episodes is thought to represent the Eddington luminosity of the
neutron star $L_{Edd}$, we can use the maximum peak flux from all of the
radius-expansion bursts, $F_{peak} = 6.3 \times 10^{-8}$ \ergcms, to
estimate the distance to the source $D$ via $L_{Edd} = 4\pi D^2
F_{peak} = 3.6 \times 10^{38}$~\ergsec (see Van Paradijs, Pennix, \&
Lewin, 1988), where we have taken the
$L_{Edd}$ corresponding to a pure helium photosphere for a 1.4~\Msun\
neutron star, and the redshift of the photosphere to be zero when 
the peak flux is measured. 
We use the largest value because some variation in the peak flux in 
radius expansion bursts 
is to be expected, as the observed Eddington luminosity decreases 
when a lower fraction of helium is present in the photosphere, when 
there is any anisotropy in the burst emission, and when the redshift of the 
photosphere at its maximum observed expansion is larger.  
Our estimate places an upper limit to the distance to \ksxrb\
of $D \approx 7$~kpc. 
The previous estimate of the distance to
\ksxrb\ was derived from the flux from Burst~1 in the same manner as
above by \citet{smb97}. The
interstellar absorption column may be variable by as much as a factor
of two, and therefore is not a useful measure of the distance to the
source (see \citep{bar98}.

The spectral parameters from three representative bursts --- a large
burst without radius expansion (Burst~3), a large burst with radius
expansion (Burst~9), and a weak burst without radius expansion (of
which we observe only one case, Burst~7) --- are displayed in
Figure~\ref{bspec}.  The apparent radius of the blackbody has been
plotted assuming that the distance to \ksxrb\ is 7~kpc. The radius
expansion bursts tend to be shorter than the bursts without radius
expansion (as indicated by $\tau$ in Table \ref{burst}).

We have indicated the spectral state during which the bursts occurred
by the numbers in Figure~\ref{cc}.  The placement of the numbers
indicates the colors of the persistent emission immediately before the
burst, and a square has been drawn around the burst number if the
burst exhibited radius expansion. All of the bursts which exhibit
radius expansion episodes (Bursts~1, 2, 8, and 9) occurred while
\ksxrb\ was on the Banana Branch, with hard color less than 0.5, while
all of the large bursts without radius expansion episodes (Bursts~3,
4, 5, and 6) occurred while \ksxrb\ was in the Island State, with hard
color greater than 0.6. The single small Burst~7 occurred on the
banana branch, which is not surprising given its small $\tau$ and
short decay time scale. This behavior was also seen in EXOSAT
observations of other sources (e.g. \citep{vdk90}).

\begin{figure}
\plotone{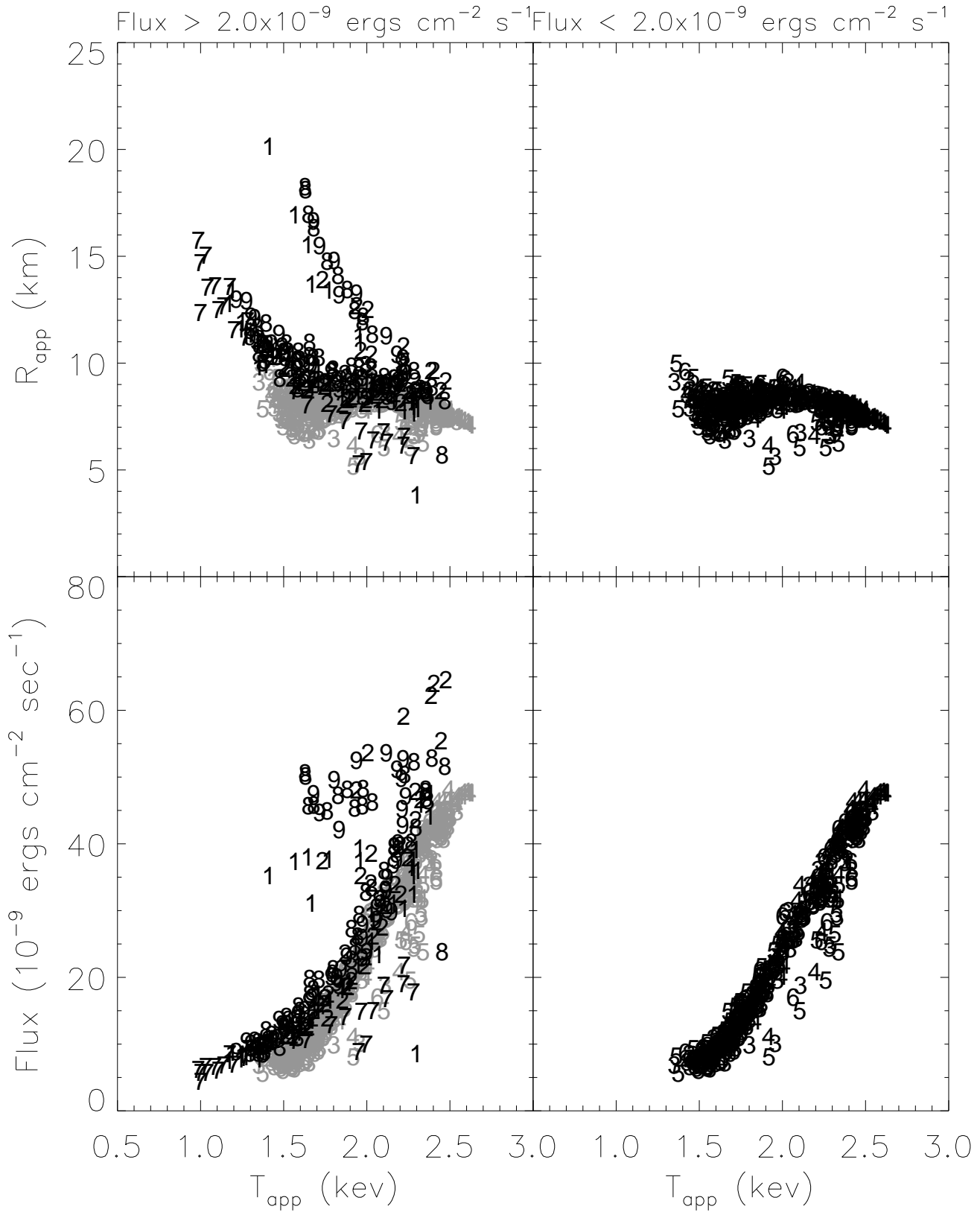}
\caption{Apparent radius (top) and bolometric flux (bottom) plotted 
as a function of apparent temperature during the bursts. The left panels 
contain fast bursts, and the right panels contain slow bursts (which are
also plotted in grey on the left panels). The numbers indicate the 
burst from which the data was taken (see Table~2). Error bars on the
plotted values are comparable to the size of the numbers used to mark 
each point.}
\label{lvsr}
\end{figure}

Following Gottwald \etal\ (1986), in Figure~\ref{lvsr} we have plotted
the apparent radius during bursts from \ksxrb\ (top panels) and the
bolometric flux in the burst (bottom panels) as a function of apparent
temperature. The numbers in the figure indicate the burst from which
the data have been taken. We have only plotted points where the burst
flux is greater than 15\% of the peak flux from the burst, because at
lower fluxes the uncertainty on the spectral parameters becomes large
(see Figure~\ref{bspec}). The panels on the left, for values of the
persistent flux greater than $2.0\times 10^{-9}$ \ergcms, represent
the bursts which occurred on the Banana Branch.  The panels on the
right, for values of the persistent flux less than $2.0\times 10^{-9}$
\ergcms, represent the bursts which occurred in the Island State.

\placefigure{lvsr}

A two-branch pattern in the $R_{app}$ vs.\ $T_{app}$ relationship is
evident in bursts from the Banana branch (top-left panel of
Figure~\ref{lvsr}), which is not present for the bursts from the
Island State (top-right panel). The right-side ($T_{app} \sim 1.5-2.5$
keV) excursion to high $R_{app} \sim 12-22$~km represents the radius
expansion phase at the start of the burst (note that Burst~7 exhibits
no radius expansion, and therefore does not follow this track), while
the excursion at lower temperatures ($T_{app} \sim 0.8-1.5$ keV,
$R_{app} \sim 10-17$~km) represents an increase in the apparent
emitting area in the tails of the fast bursts (e.g.\
Fig.~\ref{bspec}b).

It is impossible to precisely determine the time interval between
bursts and the ratio of energy released in the persistent to the
bursting luminosity ($\alpha$), because \ksxrb\ was occulted by the
Earth for about a third of every 90~minute \rxte\ orbit, and the
observed bursts occurred infrequently. We can only roughly estimate the
rate with which bursts occurred and the values of $\alpha$ during four
intervals with more than 10~ks of exposure time over the course of a
few days (see Table~\ref{obs}).  The values that follow are uncertain
by about a factor of two because of the small number of bursts
detected.  For 1997~October~28 to 29, the source was on the Banana
Branch and one burst was observed during 13.9~hours of exposure,
suggesting an approximate rate of 0.07~hr$^{-1}$.  Integrating both
the persistent (Table~\ref{obs}) and bursting flux (Table~\ref{burst})
over the exposure time during this interval, we find $\alpha \sim
640$. No bursts were observed during 13.9~hours of observations during
the interval from 1998~July~31 to August~1, also on the Banana Branch,
which is consistent with the single burst seen in the first interval.
\ksxrb\ was in the Island State for 1998~October~2 to 6, and the
approximate burst rate was $0.24$~hr$^{-1}$ with $\alpha \sim
30$. During 1999~February~22 to 27 the source was again on the Banana
Branch, and the burst rate was $\sim 0.22$~hr$^{-1}$ with $\alpha \sim
200$.  The burst rate does not seem to show any correlation with the
spectral state of the source, which suggests that some parameter other
than the accretion rate determines the frequency with which bursts
occur.  On the other hand, the values of $\alpha$ are at least a
factor of six higher when \ksxrb\ is on the Banana Branch ($\alpha
\sim 200-690$) than when it is in the Island State ($\alpha \sim 30$).
This is consistent with the previous EXOSAT phenomenology as well,
which pointed to higher $\alpha$ values when the bursts were short and
radius expansion was prevalent, and thus to more helium-dominated
bursts. 

We have attempted to determine whether bursts could have occurred
regularly during the four time intervals above by assuming that they
recur with times distributed in a Gaussian fashion about a mean time
interval.  We have determined the probability that we could have
observed bursts at the times in Table~\ref{burst} given recurrence
intervals between 0.2 and 20 hours.  Depending on the interval, we
find that either (1) no constant mean recurrence time is consistent
with the observed bursts (and lack thereof); or (2) the only
consistent recurrence times are longer than our estimate of the burst
times above, which would suggest that we were ``lucky'' to observe the
bursts that we did. An alternative possibility is that bursts from
\ksxrb\ occur in clusters, similar to the behavior of Ser~X-1
\citep{szt83}, \uxrbs{1735}{44} \citep{lew80, par88}, and
\uxrbs{1636}{53} \citep{lew87}.

\subsection{Search for burst pulsations}

In a uniform search for burst pulsations, we have produced power
density spectra (PDS) for 1~s intervals for the first 15 seconds of
the bursts (starting from the second before the burst rise), using
data from the 2--20~keV energy range binned to a time resolution of
$2^{-13}$~s.  We have searched these PDS over 5~Hz frequency intervals
centered at 262, 524, and 1048~Hz for signals at the sub-harmonic,
fundamental, and first harmonic of the signal observed by 
\citet{smb97} in the first burst. Our
detection threshold power (99\%-confidence) is 24.4 when normalized
according to the criteria of Leahy \etal\ (1983), taking into account
all of the examined bins.  We then took into account the distribution 
of noise powers in the PDS, and inverted the function of 
\citet{groth} using the algorithm in the Appendix of 
\citet{vau94} to calculate either 
1-$\sigma$ uncertainties on 
measured powers, or 99\% upper limits to the true powers when none 
was detected significantly. Assuming any signal power (or upper limit) 
is produced by a sinusoidal variation of amplitude $A$, we can invert 
the relationship 
$$<P_j> = 0.773 N_{\gamma} {{A^2}\over {2}} 
	{\rm sinc}^2 ({{\pi} \over {2}} {{\nu_j} \over {\nu_{\rm Nyq}}}),$$
where $<P_j>$ is the power in frequency bin $\nu$, and $\nu_{\rm Nyq}$
is the Nyquist frequency of the PDS, in order to find the oscillation 
amplitude \citep{lea83}. Table~\ref{osc} reports the RMS amplitudes 
of the largest detected oscillations, or upper limits on oscillations 
derived from the largest observed power, for each of the
bursts. Five signals have been detected at a frequency of 524 Hz, and no
significant signals have been detected at the sub-harmonic or first
harmonic. The signal in Burst~7 is significant with only
90\%-confidence given our broad search parameters, but since there is
a (much smaller) 0.2\% chance that the signal could be due to noise in
a search restricted to the first second of the burst rise, and since
the significance of these pulsations has been confirmed by phase
modeling (see below), we treat this signal as a detection. The RMS 
amplitude of the oscillation in Burst~7 is large only because the count 
rate in the rise of the burst is a factor of 2 smaller than the other 
bursts (compare Table~\ref{burst}).

\placetable{osc}

A brief comparison of Table~\ref{osc} and Figure~\ref{cc} reveals a 
striking discovery.
Bursts during which coherent oscillations were observed have been
marked with a diamond in Figure~\ref{cc}. Along with Table~\ref{osc},
this figure reveals that all four of the burst signals that occur
after the rise of the burst were found in bursts that exhibited radius
expansion episodes (Bursts 1, 2, 8, and 9). As a signal is also found
in the rise of Burst~7, all of the bursts seen on the Banana Branch
exhibited coherent oscillations. On the other hand, none of the bursts
in the Island State (Bursts 3, 4, 5, and 6) exhibited coherent
oscillations (to the sensitivity of our search).

\subsection{Characterization of burst pulsations}

\begin{figure}
\plotone{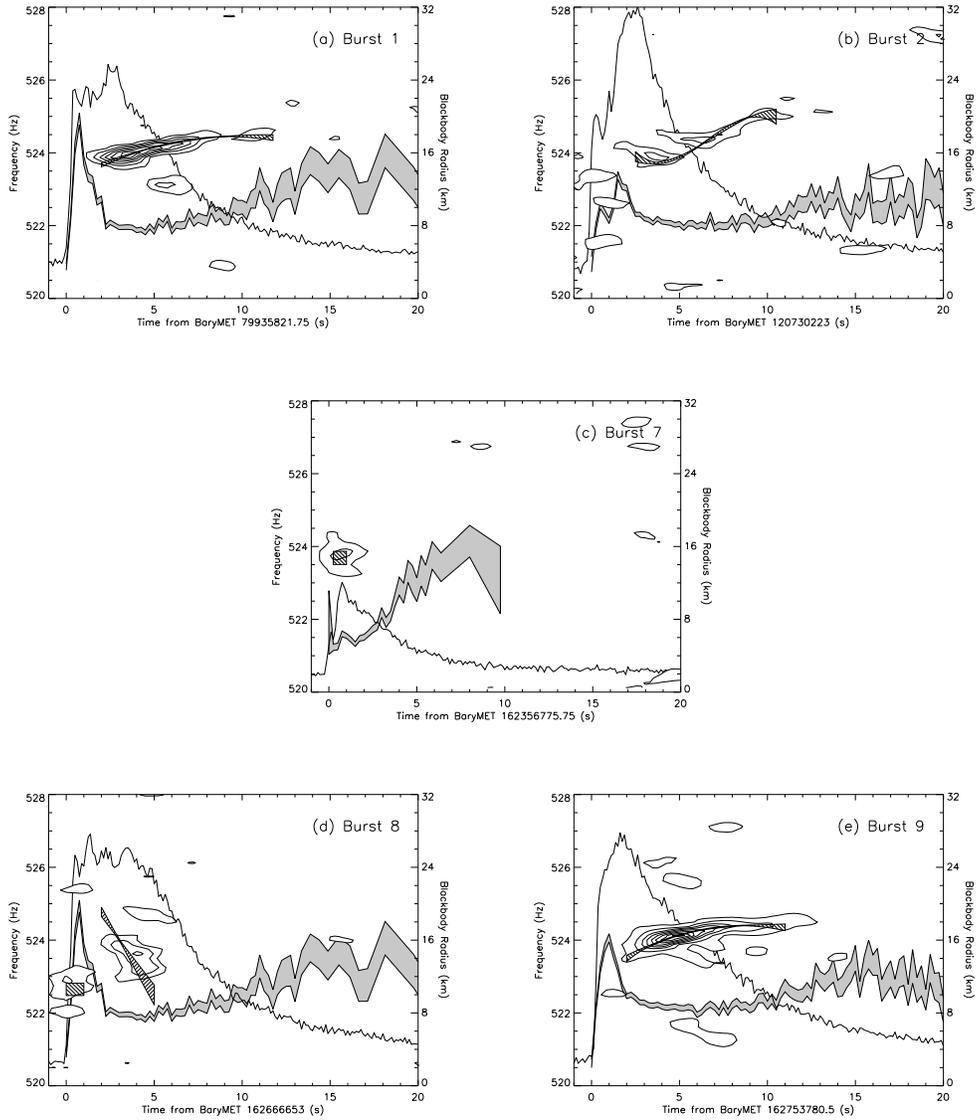}
\vspace{-1in}
\caption{Frequency evolution of the burst pulsations in the bursts
where we detect them.  Contours of the dynamic power-density spectra
are shown, with contour levels at powers of $10^{-1}$ in single-trial
probability starting at a chance occurence of 10\% (outer contour). 
The frequency evolution derived from our phase model
(approximate $\pm$1-sigma confidence region) is overplotted where
significant pulsations are detected.  Burst profiles (total PCA count
rate, on a scale from 0 to 24,000 \ctsec) and fitted blackbody radii
(at a distance of 7~kpc; $\pm$1-sigma confidence region) are also
shown.  (a) Burst~1, 14~July 1996; (b) Burst~2, 29~October 1997; (c)
Burst~7, 23~February 1999: burst spectral fits are not well
constrained past $\Delta t = 10$~s; (d) Burst~8, 26~February 1999: two
distinct pulsation episodes are detected and modeled in this burst;
(e) Burst~9, 27~February 1999: contour levels start at $10^{-1}$ and
are spaced by powers of $10^{-3}$ in single-trial probability. For
details of the construction of the dynamic power-density spectra,
please see text; phase model parameters are given in Table~4.}
\label{fev}
\end{figure}

For the bursts with detected oscillations, we have examined the
evolution of the burst pulsations by creating a dynamic power-density
spectrum (Fig.~\ref{fev}).  
For these plots we have executed 4~s FFTs at 0.25~s intervals,
oversampling the PDS by a factor of two to improve frequency
resolution (see, e.g., Middleditch (1975); Chakrabarty (1996)) and
applying a Welch window function (Press \etal\ 1992) to reduce the
side-lobes on strong powers.  For all bursts we have found
that the strengths of the oscillations were enhanced by restricting our
analysis to higher energy photons: for Burst~1 we select photons with
energies above 3.6~keV, and for the remaining bursts we select photons
with energies above 5.1~keV.  Figure~\ref{fev} shows the character of
the frequency evolution in each case.  In most cases contour levels
represent powers that are powers of 10$^{-1}$ in single-trial probability,
starting with a 10\% chance detection (where the signal is likely to 
be noise) at the outermost contour, and probablities of $10^{-2}$, $10^{-3}$, 
etc. for inner contours; in Burst~9,
however, the pulsations are significantly stronger: the first contour
is at a single-trial probability of $10^{-1}$, with further contours
spaced by powers of 10$^{-3}$.

\placefigure{fev}

For comparison purposes we also have provided, in the plots of
Fig.~\ref{fev}, the 2--60~keV light curve of the bursts from Standard1
data (running from 0 to 24,000 PCA \ctsec, full scale) and the burst
blackbody radius as derived from our spectral fits (see above; a
distance of 7~kpc is assumed).  In Bursts~1, 2, 8, and 9 the burst
pulsations commenced when the blackbody radius dropped below
\twid{10}~km; in Bursts~7 and 8 a short set of oscillations occurred in
the rise of the burst.  Finally, we have plotted the frequency
evolution as derived from our phase models (see below; approximate
plus or minus one-sigma ranges are indicated), which confirms in more
precise terms the overall picture given by the dynamic PDS.

We have observed frequency evolution in all cases, except during the
rise of Bursts~7 and 8, when the duration of the signal is short,
$<$1~s.  The form of the frequency evolution in Bursts~1 and 9 is very
similar: in both cases the frequency increases by approximately 0.5~Hz
during the course of the burst decay, relaxing upwards in a
quasi-exponential fashion. This behavior was not observed by 
\citet{smb97} because they created a single
PDS every second, providing them with a frequency resolution of only
1~Hz. The frequency evolution we see in these bursts is strongly
reminiscent of that seen in the burst sources \slowb\ \citep{str98b}
and \uxrbs{1636}{53} \citep{sm99}, although the magnitude of the
fractional frequency shift is a factor of $\sim 10$ smaller.

Burst~2 shows the most complicated behavior, starting at a frequency
of approximately 524~Hz and evolving towards 525~Hz in a sort of
S-curve.  Burst~8, in addition to showing pulsations on the burst
rise, shows a substantial frequency decrease during the burst peak,
joining the burst from \uxrbs{1636}{53} analyzed by 
\citet{str99} and the additional burst from \uxrbs{1636}{53}
analyzed by \citet{mil00} as counterexamples to the
spin-up behavior that initially seemed to characterize the burst
oscillations.

\subsection{Phase connection of burst pulsations}

In order to make a quantitative study of the form of the frequency
evolution, we have performed {\it phase connection\/} of the burst
pulsations.  The technique of phase connection is familiar from
traditional radio (Manchester \& Taylor 1977) and X-ray pulsar 
studies (e.g. \citet{dbp81}, \citet{kcs99}), and allows for a precise
characterization of the pulse train from a source even when individual
pulses are not detectable in the data.  Our phase connection procedure
works as follows.

Beginning with the dynamic PDS, we identify the interval of strongest
pulsations and their corresponding average frequency.  We take each
half-second of data from the burst, fold at the average frequency,
test for the presence of significant modulation, and if sinusoidal
modulation is present (with $>$90\% confidence), cross-correlate the
pulse profile with a sinusoid to determine the reference phase of the
pulsation at the start of that half-second interval. We use twelve
phase bins for our folded profiles; our results are not sensitive to
this choice.  The uncertainty in the reference phase is determined by
Monte Carlo (MC) simulation: we randomly perturb the count rate for
each phase bin within the range expected from Poisson counting noise
and recalculate the reference phase 100 times for each phase point;
our phase uncertainty is the standard deviation of these MC phases.

With reference phases and uncertainties in hand we connect the
phases with a best-fit polynomial phase model.  The accuracy of this
model is then confirmed by epoch folding and phase referencing (i.e.,
the procedure described above) with respect to the model: a successful
model, by accurately tracking the pulsations through their full
evolution, will produce phase residuals scattered around zero, whereas
an inaccurate model (where it evinces any modulation at all) will
produce phase residuals evenly covering the ($-$0.5, 0.5) interval
past the point where the model breaks down.  We take reference phases
to lie between $-$0.5 and 0.5.  Since the derived phase uncertainties
are $<$0.10, chances that a \twid{10}-point phase model will
serendipitously be ``successful'' by this criterion are entirely
negligible.  In some cases, the pulsations are strong enough that we
can work with 0.25~s time intervals, and for the sake of better model
constraints we do this for the pulsations during the rise of Bursts~7
and 8.

The above procedure can be iterated, as necessary, or applied to
arbitrary phase models until a satisfactory chi-squared value is
obtained.  In our analysis we typically added terms until the model
had the flexibility to accomodate the evolution seen in the dynamic
PDS and, moreover, had a reasonable chi-squared value.  Naturally, a
(more or less) smooth polynomial evolution of the frequency is not
guaranteed a priori; on the contrary, 
\citet{str99} and \citet{mil00} have shown
that piecewise-linear frequency drift models --- with occasional
sudden jumps in $\dot{\nu}$ --- describe the frequency evolution in
several bursts from \uxrbs{1636}{53} quite well.  Nevertheless we have
found polynomial models adequate for all of the \ksxrb\ bursts.  A
detailed description of our phase models is presented in
Table~\ref{model}.

\placetable{model}

In order to determine the approximate one-sigma ranges shown in the
plots of Figure~\ref{fev}, we have set each parameter of the fit (in
turn) to its one-sigma limit as shown in the Table, minimized $\chi^2$
with respect to the remaining parameters, and calculated the frequency
evolution throughout the pulsation interval, as derived from the fit.
The full range in frequency sampled by each point in time, in the
course of sampling the different fits, was saved and determined the
confidence regions that we have plotted.  Although this procedure
falls short of the full exploration of $\chi^2$ space that would be
required for exact limits, we feel that it captures the essence of the
flexibility allowed with respect to the best-fit model.  The flaws
with our approximate method are most apparent in the plot for Burst~2,
Figure~\ref{fev}(b), where the uncertainty in the frequency at
intermediate times during the evolution is certainly greater than
shown.

Inspired by the form of the frequency evolution shown in the dynamic
PDS of Bursts~1 and 9 (Fig.~\ref{fev}), and by the work of 
\citet{sm99}, we have also fit
exponential-relaxation phase models to the data from these two bursts.
Note that the final polynomial phase models for these bursts have four
terms each (quadratic frequency evolution) so that we could pursue
exponential models without sacrificing any degrees of freedom.  We
have found that the exponential-relaxation models, $\nu(t) = \nua\ -
\Delta\nu\, e^{-t/\taunu}$, provide fits as good as the polynomial
models in both cases, with chi-squared values of 10.8 for 8 degrees of
freedom (Burst~1) and 13.2 for 11 degrees of freedom (Burst~9),
respectively.  Full details of the exponential-relaxation fits are
shown in Table~\ref{expmodel}.

\placetable{expmodel}

The best fit values for the asymptotic frequencies of the burst
pulsations are 524.61$^{+0.13}_{-0.07}$ and
524.48$^{+0.05}_{-0.03}$~Hz for Bursts~1 and 9 respectively.  The
coincidence of these two values (which are consistent at the
1.5$\sigma$ level) is remarkable, with the difference being equivalent
to an orbital Doppler shift ($v\sin i$) of only 75~\kmsec.  This
near-equality of asymptotic frequencies found using an exponential 
frequency evolution model has been observed in  \slowb, \uxrbs{1636}{53}, 
and \uxrbs{1702}{429} as well \citep{str98b, sm99}, 
which suggests that the model is physically meaningful. 

\begin{figure}
\plotone{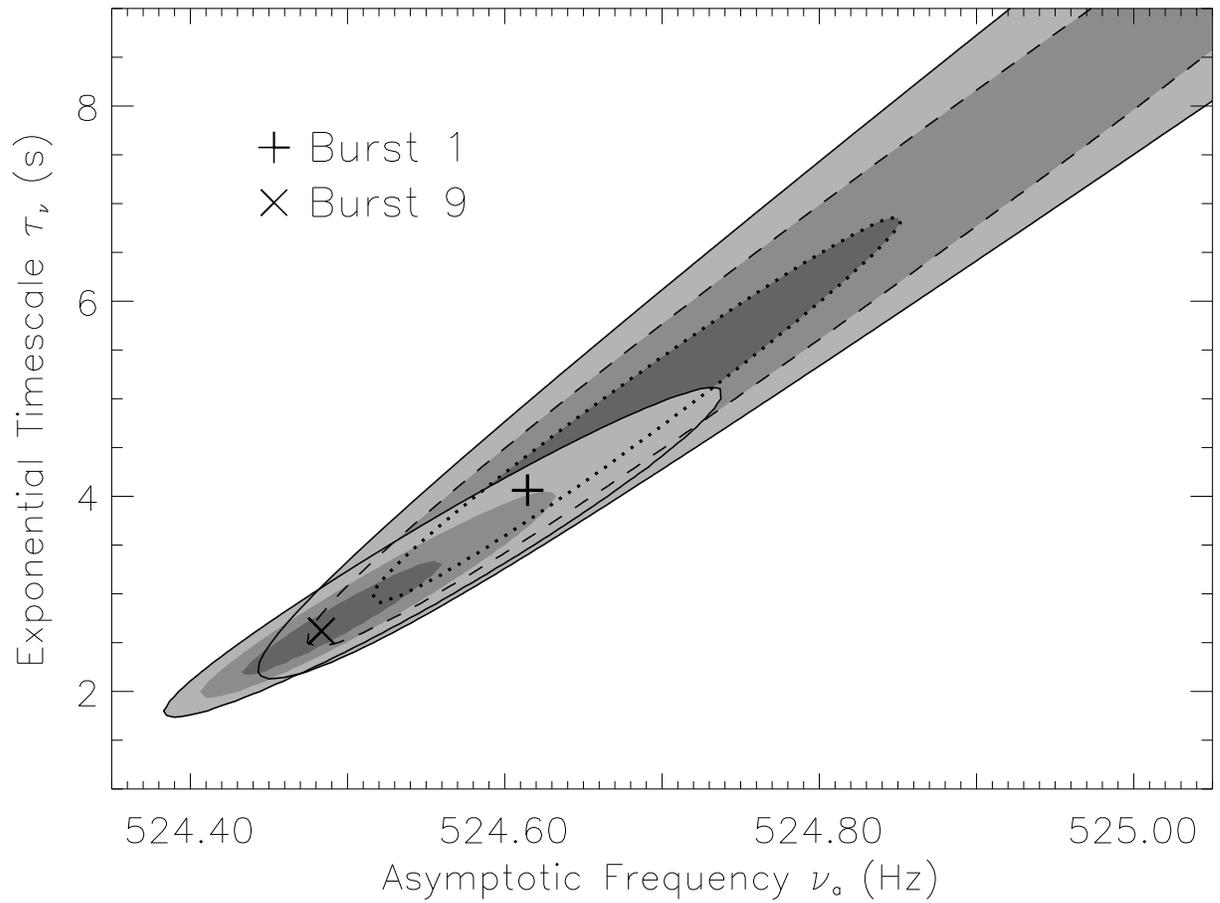}
\caption{Two-dimensional confidence regions for the \nua\ and \taunu\
parameters of the exponential fits to the frequency evolution in
Bursts 1 and 9 (see also Table~5).  Best fit values are
indicated and contours are drawn at the one-, two-, and three-sigma
confidence levels.  Parameters for the two bursts are consistent at
better than one-sigma confidence; however, if the \nua\ values for the
bursts are in fact different (e.g., due to intrinsic changes or
orbital Doppler effects) then the \taunu\ values are likely to be
different as well.  The two-sigma confidence region for Burst~1
extends to $\nua=525.34$~Hz, $\taunu=12.7$~s; the three-sigma
confidence region extends beyond this to large (\nua, \taunu).}
\label{twodreg}
\end{figure}

Proceeding to a joint investigation of the two physically
meaningful parameters of the exponential model, \nua\ and \taunu, we present in
Figure~\ref{twodreg} the two-dimensional confidence region for \nua\
and \taunu, as constrained by our exponential-relaxation model fits
for Bursts~1 and 9.  Note that values of the two other parameters,
$\Delta\nu$ and \phiz\ (the phase of the pulsations at $t=0$), are
sensitive to the exact start time of our phase model and hence depend
on our instrument sensitivity.

\placefigure{twodreg}

Examination of Figure~\ref{twodreg} reveals, first of all, that the
fit parameters for Bursts~1 and 9 are consistent at better than
1$\sigma$, as the 1$\sigma$ confidence regions for the two bursts
overlap near $\nua=524.54$~Hz, $\taunu=3$~s.

Second, the figure shows that a \nua-\taunu\ degeneracy weakens our
constraints at large (\nua, \taunu).  The sense of this degeneracy is
that increasing both \nua\ and \taunu\ moves the final approach to a
constant frequency out to later times, and so will be allowed to the
extent that the frequency is still increasing when the oscillations
disappear.  Since the pulsations in Burst~1 are weaker than those in
Burst~9, especially at late times, the problem is more acute for this
burst.  Nevertheless, in examining the plot we may draw some
conclusions.  First, if the \nua\ values for the two bursts are in
fact equal (or nearly equal), then the \taunu\ values for the bursts
must also be nearly equal, within \twid{10\%}.  Alternatively, if the
\nua\ values for the two bursts are in fact different, whether because
of an intrinsic change in the pulse frequency or because of orbital
Doppler effects, then the \taunu\ values for the bursts are also
likely to be different.

\subsection{Application of phase models}

The phase models in Table~\ref{model} demonstrate full coherence of
the burst pulsations from \ksxrb\ over the duration of the intervals
we model. The longest of these intervals are: for Burst~1, 9.75~s or
5111~cycles; for Burst~2, 8~s or 4194~cycles; and for Burst~9, 9.0~s
or 4718~cycles.  With successful phase models for the burst pulsations
we may address a number of issues.

First, we can make sensitive tests for the presence of harmonics and
sub-harmonics of the burst pulsations.  We searched first for
significant modulation at twice the main burst frequency.  By
multiplying the phases derived from our phase model by two, and
integrating through the burst pulsation intervals indicated in
Table~\ref{model}, we have found evidence at the 2.1$\sigma$ level for
a modulation at the first harmonic of the 524~Hz pulsations.  The
average power of this modulation, over all the intervals we have
modeled, is 1.9$^{+1.8}_{-0.9}$\% that of the main signal. We then
searched for and found no evidence for modulation at the second and
third harmonics of the main signal; 2$\sigma$ upper limits are 2.3\%
and 1.7\% the power of the main signal, respectively.

We then searched for a sub-harmonic of the main signal by multiplying
our derived phases by 0.5; as \citet{mil99} points
out, we must also account for the possibility of half-phase offsets at
the points of connection between different burst intervals.  Since the
signals from Bursts~1 and 9 are significantly stronger than the
signals in the other bursts, it is possible that the increased number
of trials required to connect all of the intervals could offset the
increased sensitivity provided by coherent phase connection of those
intervals.  Therefore we have performed this search in two stages.  In
the first stage, we tested for a sub-harmonic in the coherent
connection of Bursts~1 and 9 only (two trials): we did not detect
significant ($>$95\%-confidence) modulation in either connection, and
can place a 2$\sigma$ upper limit on the strength of a sub-harmonic in
these bursts, relative to the main signal, of 5.5\%.  Note that the
sub-harmonic detected by \citet{mil99} in the
burst-rise pulsations from \uxrbs{1636}{53} was on average 19\% the
power of the main signal.

In the second stage of the sub-harmonic search we connected all six of
the burst pulsation intervals together, accounting for all possible
combinations of half-phase offsets (32 trials): we did not detect
significant modulation in any connection, and place a two-sigma upper
limit on the average power of a sub-harmonic across all these
intervals of 3.9\% that of the main signal.  At the same time we
tested for the presence of modulation at 1.5 times the main burst
frequency, and can place a two-sigma upper limit on the average
power at that frequency of 4.4\% that of the main signal.

The last two points of the phase model for Burst~2 may be used to
derive a ``peak frequency'' for the pulsations in this burst of
525.08$\pm$0.18~Hz.  Although this frequency is not as well
constrained as the asymptotic frequencies of Bursts~1 and 9, it is
higher than the Burst~9 \nua\ at a 3.2$\sigma$ level of confidence,
and requires a relative Doppler shift of 340$\pm$100~\kmsec\ to
explain, if the \nua\ we measure in fact corresponds to the spin
frequency of the neutron star -- since the burning layer may expand to
produce lower frequencies, but cannot ``contract'' to produce higher
ones.  

\begin{figure}
\plotone{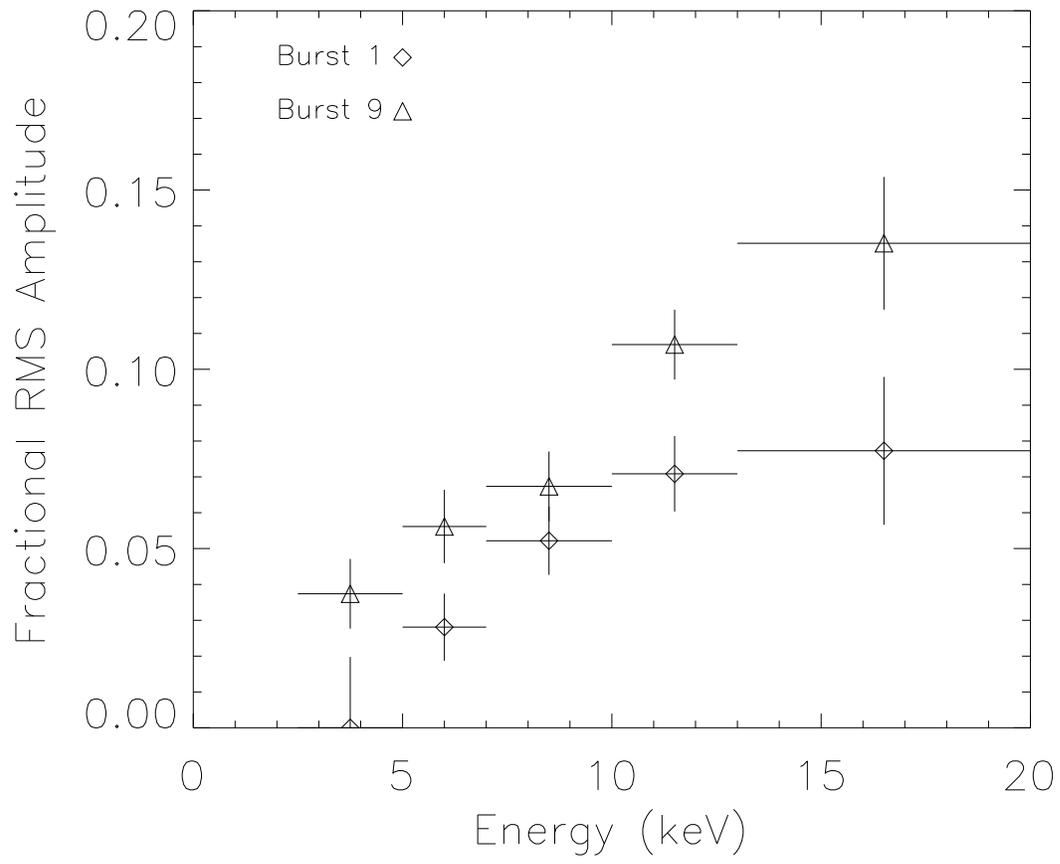}
\vspace{-1in}
\caption{Energy spectrum of the oscillations, derived from a sinusoidal fit
to the pulse profile about the best-fit polynomial phase evolution.}
\label{oscvse}
\end{figure}

For the rest of our analysis we concentrate on the pulsations from
Bursts~1 and 9, which have the highest signal-to-noise.  We have
folded data with $2^{-13}$~s time resolution in five energy channels
(starting at 3, 5, 7, 10, and 13 keV) from Bursts~1 and 9 about the
polynomial phase models. Fitting a sinusoid to the folded profile in
each of five energy bands, we measure the strength of the oscillation
and the phase offset in each band. The reduced chi-squared values for
the sinusoidal fits were less than 1.3 for all energy bands. We have
found that the RMS amplitude of the burst oscillation increases
monotonically with energy for both bursts; the sense of this increase
is shown in Figure~\ref{oscvse}.

\placefigure{oscvse}

The measured phase of the oscillations in different energy bands
provides marginal evidence for a soft phase lag in the 7--9~keV and
10--13~keV bands relative to the 3--5 keV band in Burst~1
($-$1.5$\pm$0.3 and $-$1.2$\pm$0.33 radians, respectively), and none
in Burst~9. Computing the Fourier phase lag for a signal at 524~Hz
(with 1~Hz frequency resolution) without adjusting for the polynomial
phase model, we have found a marginally significant hard lag of
0.4$\pm$0.2 radians between the 5--7 keV and 7--25 keV bands in
Burst~9, and none in Burst~1. Given these ambiguous detections, we do
not feel our measurements are significant evidence of a phase lag in
the pulsed signal.

\begin{figure}
\plotone{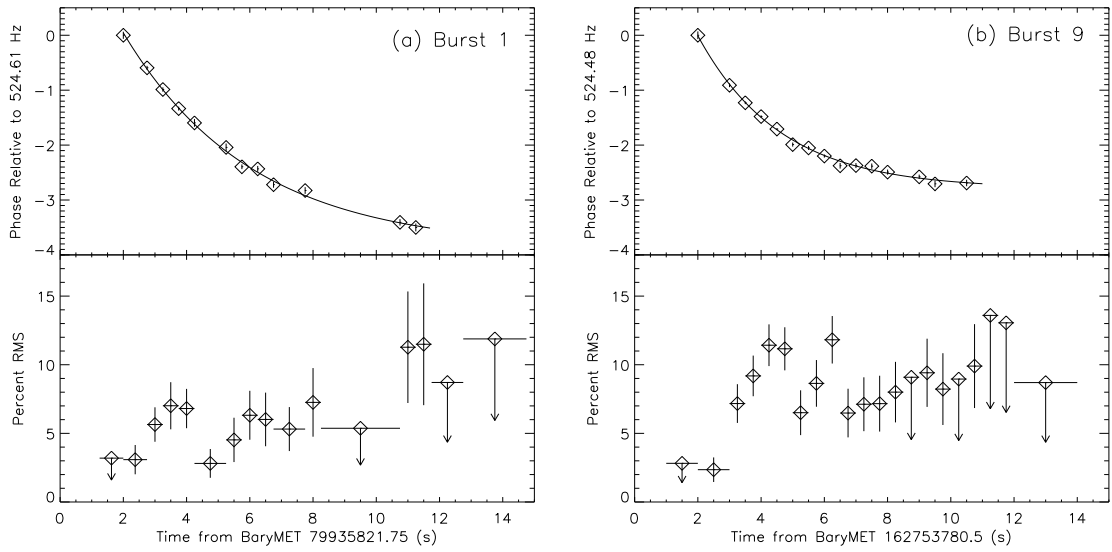}
\vspace{-1.5in}
\caption{Evolution of the pulsations in Bursts 1 and 9, as determined
from our phase models.  (a) Burst~1: (upper) Evolution of the phase of
the pulsations, relative to a constant frequency of 524.61~Hz, with
exponential-relaxation model superposed; (lower) Strength of the
pulsations throughout the burst, relative to the total 3.6--60~keV
burst flux.  (b) Burst~9: (upper) Evolution of the phase of the
pulsations relative to a constant frequency of 524.48~Hz, with
exponential-relaxation model superposed; (lower) Strength of the
pulsations throughout the burst, relative to the 5.1--60~keV burst
flux.  Upper limits are 2$\sigma$.  }
\label{expev}
\end{figure}

Finally, we have been able to accurately characterize the evolution of
the strength of the pulsations over the course of the bursts
(Fig.~\ref{expev}).  Without an accurate phase model, the strength of
the pulsations would otherwise be very difficult to measure when they
first appear and their frequency is changing the fastest.  It appears
that the strength of the burst pulsations (relative to the burst flux)
in Burst~1 are consistent with a \twid{7\%}~RMS modulation for most of
the burst, rising to a maximum of \twid{10\%} before disappearing at
the end; in Burst~9, the data show a significant increase at the start
of the pulsations, and are then consistent with a more or less
constant 8\%~RMS modulation for the duration of the pulsations.  In
both bursts, the pulsations are initially detected at a lower level of
\twid{3\%}~RMS.  

\placefigure{expev}

We have calculated the strength of the oscillations in the other
bursts where they are detected, although we do not present the results
as figures: the pulsations in Burst~2 increase from $3\pm
1$\%~RMS to $11\pm 3$\%~RMS over the course of the interval when
pulsations are detected; those in Burst~7 are consistent with
$10\pm1$\%~RMS throughout the 0.75~s of detected pulsations; those in
Burst~8 have $11 \pm 4$\%~RMS at start, decreasing to $4\pm 1$\%
before disappearing.

Figure~\ref{expev} also shows our phase model fits and residuals for
Bursts~1 and 9, presented in the form of a ``phase loss'' plot ---
with the running phase compared to that for a constant-frequency
pulsation at the fitted asymptotic frequency \nua.  According to our
models, the burst pulsations lose 3.9$^{+3.4}_{-1.1}$ cycles compared
to the constant \nua\ in Burst~1, and 2.8$^{+0.7}_{-0.4}$ cycles
compared to the (different) constant \nua\ in Burst~9.  Note that due
to the \nua-\taunu\ degeneracy in the fits, these values are not well
constrained from above (especially for Burst~1): our two-sigma upper
limits on the total phase loss are 19.4 and 4.3 cycles for Bursts~1
and 9, respectively.

\section{Discussion}

Our results allow us to address a number of important aspects of the
phenomenology of type~I X-ray bursts and their burst pulsations.

\subsection{Two populations of type I X-ray bursts}

The nine bursts that we have observed from \ksxrb\ divide into
two populations, as shown in Table~\ref{burst} and Figure~\ref{cc}.
When the source is in the Island State and the persistent flux is low,
bursts have relatively broad maxima (as measured by the time until the
start of the burst decay), long decay time scales ($\tau$),
large fluences, and no evidence for radius expansion (Bursts~3, 4, 5,
and 6); following \citet{got86} we call these
{\it slow bursts.}  When the source is on the Banana Branch and the
persistent flux is high, the peaks of bursts are sharper, the bursts
are shorter, radius expansion episodes are present, and burst fluences
are smaller (Bursts~1, 2, 8 and 9); we call these {\it fast bursts.}
Burst~7 is particularly weak, but still exhibits the short decay,
narrow maximum, and small $\tau$ characteristic of fast bursts.
When we compare the ratio of the energy released in the persistent 
luminosity to that in type~I X-ray bursts, $\alpha$, we find yet another 
distinction: the slow bursts which occur in the Island State correspond
to an $\alpha \sim 30$, while the fast bursts which occur on the Banana
Branch correspond to higher $\alpha > 200$. 

The decrease in burst duration with increasing accretion rate is
consistent with the global relationship which 
\citet{ppl88} reported for their sample of 10 LMXBs.
Note however that they found burst properties to span a
continuum between the slow and fast populations in a sample of many burst
sources, and that \ksxrb\ itself may display a continuum of bursting 
properties should the source be observed for a longer period of time at 
an intermediate value of the accretion rate as inferred from the 
color-color diagram (Figure~\ref{cc}).  The work of Hanawa,
Fujimoto, \& Miyaji (1981) (see also \citet{ppl88}, \citet{got86})
suggests that the fast bursts (and values of $\alpha \gtrsim 100$) result from 
the burning of helium-rich material. The association between fast 
bursts and radius expansion has
previously been noted by several authors (see, e.g., \citet{got86} for
EXO~0748$-$676 and \citet{lew87} for \uxrbs{1636}{53}), and indicates
that the nuclear energy of these bursts is released so rapidly
that the Eddington luminosity is exceeded, causing substantial
expansion of the neutron star's photosphere.

Hanawa, Fujimoto, \& Miyaji (1981) predict that helium-rich bursts
should occur when the local accretion rate in the area responsible for
the burst is low (Case~2 of Section~1, above).  Although this may seem
to be contradicted by our finding that fast bursts occur at relatively
high accretion rates, the paradox can be resolved if accretion takes 
place over a larger fraction of the neutron star when the total
accretion rate is high, such that the accretion rate per unit area is
reduced (Bildsten 2000). 
 Evidence for this hypothesis may be found in spectral fits to
the burst emission.  \citet{got86} noted that
the fitted blackbody radii during the tails of fast bursts from
EXO~0748$-$676 were systematically larger than the radii for the slow
bursts and Bildsten (2000) showed that this is adequate to 
change the local accretion rate. 
Similarly, we find that the apparent radii in the tails of
fast bursts from \ksxrb\ are systematically larger than those of slow
bursts (Figure \ref{lvsr}); this behavior can be seen in the spectral
evolution of Bursts~3 and 9 in Figure~\ref{bspec}. If this difference
in apparent blackbody radii reflects a physical increase in the area
of the burst emission, that might explain why the properties of bursts
observed at high total accretion rates are those expected for bursts
occurring at a low local accretion rate. 

The behavior of Burst~7 runs somewhat counter to this trend: although
this burst has significantly less fluence than the other fast bursts,
it exhibits a larger increase in blackbody radius during the tail of
the burst.  If a burst only ignites when sufficient material per unit
area is accreted to produce the critical density and temperature, then
larger burst areas should correspond reliably to larger burst
fluences.  On the other hand, if a burst occurred during the Earth
occultation 25--75 minutes prior to Burst~7, then the observed burst
would be similar to other weak bursts with short waiting times and
relatively large apparent radii from \uxrbs{1705}{44} \citep{lan87}
and EXO 0748$-$676 \citep{got86}. These weak bursts may be triggered
by unburnt nuclear fuel from the previous burst \citep{lew87, fuj87}.

It is also possible that there is a systematic difference in the
neutron star atmosphere or the accretion flow which causes the
spectrum of bursts on the Banana Branch to systematically deviate from
a blackbody due to electron scattering. London, Taam \& Howard (1984,
1986) demonstrated that at luminosities close to the Eddington value,
the ratio $\xi=\Tapp/\Teff$ of the apparent temperature derived from a
blackbody fit to the effective temperature of the emission region
increases as \Teff\ increases. Since $\Rapp \propto F^{1/2}\Tapp^{-2}
= F^{1/2}(\xi \Teff)^{-2}$, and for a constant physical emission area
$F \propto \Teff^4$, we have $\Rapp \propto \xi^{-2}$. Thus, as the
burst cools, \Teff\ and $\xi$ decrease together, and \Rapp\ will
appear to increase (see also \citet{szt85}).  If the anti-correlation
between \Rapp\ and \Tapp\ in Figure~\ref{lvsr} is indeed due to
electron scattering, then the trend is not a probe of accretion onto
the neutron star surface, but indicates that the atmosphere or
immediate environment of the neutron star is fundamentally different
in the Banana and Island states.  However, the results of London
\etal\ were originally considered important in explaining apparent
super-Eddington temperatures in the peaks of radius-expansion bursts,
and it is not clear whether the theory applies to the cooling tails of
bursts, when the temperature is nearly a factor of two
smaller. London, Taam, \& Howard (1984, 1986) find that the opposite
correlation between $\xi$ and \Teff\ holds at low \Teff.

\subsection{Fast bursts show burst pulsations}

It has previously been observed that coherent oscillations are not
present in every burst from \uxrbs{1728}{34} and \uxrbs{1702}{43}
\citep{mss99}, although no systematic study has been previously carried
out to determine whether the properties of the bursts or persistent emission
from those sources are correlated with the presence of 
detectable coherent oscillations.  We find that in \ksxrb\ 
this differentiation is
strictly consistent with the two burst populations we identify: all of
the fast bursts exhibit near-coherent oscillations and none of the
slow bursts do (to limits of \twid{10}\% RMS in any single second; see
Table~\ref{burst}). However, it is worth differentiating the
oscillations found in the rise of the bursts from those occurring in
the burst tails, as it is more difficult to understand how the latter
form.  Only two instances of coherent oscillations in the
burst rise are observed, so the question remains as to why some of the
fast bursts exhibit burst-rise pulsations and others do not. Oscillations 
in the rise of bursts are not tied to radius expansion episodes, 
as the oscillations are observed in Burst~7 from \ksxrb\, and in several
bursts without radius expansion in \uxrbs{1728}{34} \citep{szs97}. On the
other hand, all of the oscillations which are observed in the tails of
the bursts appear immediately after the contraction phase of a radius
expansion episode (as first noted by \citep{smb97}; no counter-example
of coherent oscillations in the tail of a burst without radius expansion 
has been reported among the other sources to our knowledge).  This suggests
that the radius expansion episode leaves behind an anisotropy in the
neutron star photosphere that resembles a rotating hot spot.  In any
case, the general association between fast bursts and coherent
oscillations suggests that the process which produces the oscillations
is related to the particularly intense helium-rich burning in these
bursts. 

\subsection{Burst pulsations and the spin frequency}

Our analysis of the burst pulsations raises some intriguing
questions about the spin-frequency interpretation of the burst
pulsations in \ksxrb. The best-fit values of the asymptotic burst
pulse frequency \nua\ for Bursts~1 and 9 are 524.61$^{+0.13}_{-0.07}$
and 524.48$^{+0.05}_{-0.03}$~Hz respectively, are consistent at the
1.5$\sigma$ level. The 0.13~Hz frequency difference could be produced
by an orbital Doppler shift ($v\sin i$) of only 75~\kmsec, well within
the range of typical LMXB orbital parameters. However, the highest
frequency of the oscillations in Burst~2 is greater than the Burst~9
\nua\ at a 3.2$\sigma$ level of confidence, and requires a relative
Doppler shift of 340$\pm$100~\kmsec\ if the \nua\ we measure in
Bursts~1 and 9 in fact reflects the spin frequency of the neutron
star.  A 300~\kmsec\ Doppler shift is not inconceivable for an LMXB,
but (assuming an inclination of $60^{\circ}$ as suggested by the lack 
of eclipses or dips, and a 1.4 \Msun\ neutron star with a 1 \Msun\ companion)
would suggest a 1~h orbit for the most probable viewing geometry of the 
bursts--- placing \ksxrb\ in the shortest \twid{10\%} of
the LMXB orbital period distribution \citep{wnp95}.
Unfortunately no optical companion to \ksxrb\ has
been found to date \citep{bar98}, and no orbital period is known for
the binary. More information about the system, or a greatly expanded
sample of bursts, will be necessary before the binary parameters can
be constrained.

The fractional change of the oscillation frequency during Bursts~1, 2,
and 9 (Figure~\ref{fev}) is within a factor of a few of the changes
observed in \slowb\, \uxrbs{1702}{43}, and \aqlx \citep{sm99},
\citep{zha98}.  Under the assumptions of the hot spot model, the
fractional change in the frequency is directly proportional to the
change in the radius at which the hot spot is observed. When angular
momentum is conserved, $\nu r^2 \approx (\nu - \Delta\nu)(r + \Delta
r)^2$, which for small $\Delta\nu$ and $\Delta r$ gives 
$\Delta\nu / \nu \approx 2 \Delta r / r$.  For \ksxrb\ a 1~Hz
frequency drift corresponds to a $\Delta r / r$ of $\sim 2\times10^{-3}$, or
about 20~m for a neutron star with a 10~km radius.

Comparing the various time and length scales which we can infer for
the bursts provides some insight into the origin of the oscillations.
The change in the radius of the hot spot during the tails of the
bursts ($\sim 20$~m) is thought to just arise from the hydrostatic
expansion of the nuclear burning layer \citep{szs97}.  However, the
time for the nuclear energy released at the burning layer to diffuse
through the neutron star atmosphere to the photosphere during a burst
can also be as large as several seconds \citep{hf84}. If the hot spot
is moving with respect to the surface of the neutron star (and the
atmosphere) more quickly than the signal from the hot spot can escape
the atmosphere, the signal will be unobservable.

Cumming \& Bildsten (2000) have shown that the smaller opacity of a
helium rich shell reduces the thermal diffusion time through a helium
rich atmosphere.  In addition, they found that the time for the hot spot to
drift relative to the surface of the neutron star (estimated from the
inverse of the predicted frequency change) increases when more helium
is present, because the higher mean molecular weight of helium
prevents the burning layer from expanding as far. The combination of
the two effects implies that, independent of the cause of the hotspot,
the signal is more likely to be observed during a pure helium burst.

In Burst~8 we observe an episode of spin-down during the burst peak
that is roughly comparable in duration and magnitude to the spin-down
episode observed during \citet{str99} Burst~A
from \uxrbs{1636}{53} (also discussed in \citet{mil00});
see Figure~\ref{fev}(d).  In this case, however, the spin-down is not
accompanied by any evidence for late-time radius expansion, as was
observed for Strohmayer's Burst~A; on the contrary, the entire
spin-down episode occurs after the photospheric radius has contracted
and settled to its minimum value of 8~km.  This observation presents a
challenge to the spin-frequency interpretation.  Explaining the
spin-down during Burst~8 under this model seems to require postulating
a relatively consistent \twid{7}~m~s$^{-1}$ expansion of the burning
layer over the \twid{3}~s interval of the pulsations, without any
corresponding spectral evidence for this expansion.  Naturally, a
small expansion like this would not be seen directly in the blackbody
radius of the spectral fits.  However, some additional input of
energy, relative to the more typical spin-up seen in Bursts~1, 2, and
9, is presumably required to power this expansion, and we see no
evidence of this.

The amplitude of the burst oscillations we observe increases with
photon energy (Figure~\ref{oscvse}) in a manner consistent with
general-relativistic models of hot spots on the surface of a rotating
neutron star \citep{ml98}. These models allow larger oscillation
amplitudes (20--80\% for two hot spots, as the photon energy increases
from 0--20~keV) than we observe, but this can be accounted for by
allowing the hot spots on \ksxrb\ to have a finite area on the neutron
star surface, which will dampen the modulation (the intent of Miller
\& Lamb 1998 was to calculate upper limits on the pulse amplitudes,
so point-source hot spots were assumed).

Phase connection of the pulsations in the bursts from \ksxrb\ allows
us to make sensitive tests for the presence of harmonics and
sub-harmonics of the main signal.  We find marginal evidence
(2.1$\sigma$ level) for modulation at the first harmonic of the main
524~Hz signal with an average power 1.9$^{+1.8}_{-0.9}$\% that of
the main signal, and no evidence ($>$95\% confidence) for modulation
at frequencies of 1/2 or 3/2 the frequency of the main signal; our
2$\sigma$ upper limits for the power of these components are 5.5\%
and 4.4\% the power of the main signal, respectively.  The
sub-harmonic modulation in \uxrbs{1636}{53} found during oscillations
in the burst rise has an average strength 19\% that of the main signal
\citep{mil99}, and is compelling evidence that the spin frequency of
that neutron star is in fact half of the strongest burst
frequency. This would be expected in beat frequency models, since
$\nub\approx 2\nud$. The strongest observed oscillations in \ksxrb\
are in the burst tail, so it is possible that the small upper limit on
the sub-harmonic content observed from this source (by a factor of
\twid{3}) reflects greater symmetry in the hot spots present in the
burst tail.  However, the evidence we find for modulation at the first
harmonic of 524~Hz suggests that it is more likely that $\nus =
524$~Hz in \ksxrb, since \twid{2}\% harmonic content might easily be
produced by a pulse which deviates slightly from a sinusoidal profile.
Should observations of further bursts from \ksxrb\ strengthen the
evidence for a harmonic, the case for $\nus=\nub$ in this source will
be strengthened, and the beat frequency interpretation cast into
doubt.

\section{Conclusions}


We have observed nine type~I X-ray bursts from \ksxrb\ with the \rxte\
PCA. We find that the bursts can be separated into two categories 
(Section~2.2, 2.3):
fast bursts occur on the Banana Branch when the accretion rate is
high, and generally exhibit short decay time scales, high peak flux,
radius expansion episodes, and coherent oscillations. Slow bursts
occur in the Island State at lower accretion rates, and have lower
peak fluxes, higher fluences, and longer decay time scales, but
exhibit no evidence of either radius expansion or coherent
oscillations.  The fast and slow bursts may occur when the nuclear
fuel is rich in helium and hydrogen respectively (Sections~1.1 and 3.1).

We have analyzed the frequency evolution of the pulsations during the
bursts by applying the technique of pulsar phase connection, which
allows us to ``count pulses'' and thereby characterize the evolution
uniquely (Section~2.5).  
We find that the frequency evolution during the peak and
decay of the bursts exhibits a variety of behaviors, including a sharp
spin-down in one burst (Burst~8) which challenges the attribution of
the frequency evolution during bursts to expansion or contraction of
an expanded burning layer.  The frequency evolution during Bursts~1
and 9, separated by 2.6~years, extends for over 9~s in both cases and
is consistent with nearly the same exponential-relaxation model, with
a best-fit frequency difference of 0.13$\pm$0.09~Hz. The evolution of
the oscillations in Burst~2 is more complicated, and the maximum
frequency of the oscillation is greater than those modeled with
exponentials by 0.6$\pm$0.2~Hz.

Phase connection of the burst pulsations allows us to make several
precision tests of their properties (Section~2.6).  
We demonstrate the coherence of
these pulsations (after accounting for their gross frequency
evolution) for $\gtrsim$5000~cycles.  We find that the pulsations are
spectrally hard in comparison to the burst emission, with the strength
of the pulsations increasing monotonically with photon energy.
Coherently summing the signals from all the bursts with detected
pulsations, we find marginal (2.1$\sigma$) evidence for modulation at
the first harmonic of the main pulse frequency at 524~Hz, with an
average power equivalent to 1.9$^{+1.8}_{-0.9}$\% that of the main
signal.  In combination with our upper limits on modulation at 1/2 or
3/2 the main pulse frequency (5.5\% and 4.3\% the power of the main
signal, respectively), this suggests that the spin frequency in
\ksxrb\ is more likely to be \twid{524~Hz} than half that, as the
measured frequency difference of the kHz QPOs ($\nud\approx 260$~Hz)
might otherwise suggest.  

\acknowledgments

We thank Don A. Smith for providing us with the data from his \rxte\
proposals for \ksxrb, and Deepto Chakrabarty for pointing us towards
phase connection as the best way to model the frequency evolution
during bursts (and teaching us how to do it).  
This research was partially supported by NASA via
grants NAG 5-8658 and NAGW-4517 and by the National Science Foundation
under Grant No. PHY94-07194. L. B. is a Cottrell
Scholar of the Research Corporation.

\clearpage

\begin{deluxetable}{lcccccccc}
\scriptsize
\tablecolumns{9}
\tablewidth{0pc}
\tablecaption{PCA Observations of \ksxrb\label{obs}}
\tablehead{
\colhead{Observation ID} & \colhead{Date} & \colhead{Start} & 
\colhead{End} & \colhead{Exposure} & \colhead{Count} 
& \colhead{Flux} & \colhead{Soft} & \colhead{Hard} \\
\colhead{} & \colhead{UT} & \colhead{Time} & \colhead{Time} 
& \colhead{Time (s)} & \colhead{Rate\tablenotemark{a}} 
& \colhead{($10^{-9}$)\tablenotemark{b}}
& \colhead{Color} & \colhead{Color}
}
\startdata
10416-01-01-00 & 1996 Jul 14 & 02:32:31 & 07:05:31 & 9960 & 552 & 6.38 & 1.63 & 0.40  \\ 
10416-01-02-00 & 1996 Aug 01 & 16:53:56 & 20:51:56 & 9900 & 344 & 2.74 & 1.50 & 0.39  \\ 
10416-01-03-00 & 1996 Aug 31 & 17:40:17 & 19:44:16 & 3540 & 405 & 4.73 & 1.54 & 0.38  \smstrut \\ 
20085-01-01-01 & 1997 Oct 28 & 22:12:38 & 03:41:39 & 13080 & 404 & 4.69 & 1.62 & 0.40  \\ 
20085-01-01-02 & 1997 Oct 29 & 06:22:45 & 08:28:45 & 4320 & 467 & 5.37 & 1.65 & 0.41 \\ 
20085-01-01-03 & 1997 Oct 29 & 18:00:34 & 20:00:33 & 4320 & 489 & 5.63 & 1.62 & 0.43 \\ 
20085-01-01-00 & 1997 Oct 29 & 21:07:33 & 11:39:33 & 28260 & 505 & 5.85 & 1.71 & 0.46 \smstrut \\ 
30061-01-01-00 & 1998 Jul 31 & 11:26:08 & 15:17:07 & 9000 & 377 & 4.36 & 1.65 & 0.40  \\ 
30061-01-01-01 & 1998 Aug 01 & 01:49:06 & 04:09:07 & 6480 & 381 & 4.42 & 1.53 & 0.42 \\ 
30061-01-01-02 & 1998 Aug 01 & 06:47:02 & 17:18:02 & 22980 & 369 & 4.28 & 1.63 & 0.41 \\ 
30061-01-01-03 & 1998 Aug 01 & 23:04:02 & 02:54:02 & 9480 & 369 & 4.28 & 1.64 & 0.41 \\ 
30061-01-01-04 & 1998 Aug 02 & 08:32:00 & 09:07:56 & 2160 & 378 & 4.36 & 1.65 & 0.40 \smstrut \\ 
30061-01-02-00 & 1998 Oct 02 & 03:26:02 & 08:40:02 & 11040 & 106 & 1.26 & 1.73 & 0.72 \\ 
30061-01-02-01 & 1998 Oct 02 & 13:01:58 & 14:01:58 & 3600 & 118 & 1.41 & 1.74 & 0.73 \\ 
30061-01-02-02 & 1998 Oct 03 & 01:49:53 & 04:11:54 & 5100 & 103 & 1.23 & 1.73 & 0.71 \\ 
30061-01-02-03 & 1998 Oct 03 & 05:44:53 & 09:14:53 & 7800 & 109 & 1.30 & 1.61 & 0.71 \\ 
30061-01-02-04 & 1998 Oct 04 & 05:35:44 & 10:49:44 & 11880 & 107 & 1.29 & 1.69 & 0.71 \\ 
30061-01-02-05 & 1998 Oct 05 & 16:13:32 & 17:12:33 & 3540 & 108 & 1.30 & 1.74 & 0.73 \\ 
30061-01-02-06 & 1998 Oct 06 & 01:49:29 & 09:11:29 & 15360 & 115 & 1.37 & 1.73 & 0.72 \\ 
30061-01-02-07 & 1998 Oct 08 & 10:12:10 & 10:47:10 & 2100 & 128 & 1.55 & 1.58 & 0.61 \smstrut \\ 
30061-01-03-00 & 1999 Feb 22 & 18:23:17 & 20:59:17 & 7380 & 271 & 3.22 & 1.46 & 0.42 \\ 
40409-01-01-00 & 1999 Feb 22 & 23:46:17 & 05:00:17 & 9300 & 266 & 3.16 & 1.56 & 0.42 \\ 
30061-01-03-01 & 1999 Feb 25 & 21:35:42 & 23:57:42 & 6060 & 322 & 3.76 & 1.62 & 0.41 \\ 
30061-01-03-02 & 1999 Feb 25 & 23:59:42 & 00:12:42 & 780 & 313 & 3.65 & 1.62 & 0.41 \\ 
30061-01-04-00 & 1999 Feb 26 & 15:08:47 & 20:58:48 & 13380 & 390 & 4.55 & 1.69 & 0.45 \\ 
30061-01-04-01 & 1999 Feb 27 & 11:56:54 & 14:23:55 & 6840 & 389 & 4.51 & 1.61 & 0.44 \\ 
30061-01-04-02 & 1999 Feb 27 & 15:08:56 & 17:35:56 & 6120 & 380 & 4.38 & 1.69 & 0.42 \\ 
\enddata
\tablenotetext{a}{Counts per second per PCU}
\tablenotetext{b}{\ergcms between 2--18 keV. Uncertainty is approximately
$0.01\times 10^{-9}$ \ergcms for all entries.}
\end{deluxetable}

\begin{deluxetable}{clccccccc}
\scriptsize
\tablecolumns{9}
\tablewidth{0pc}
\tablecaption{Bursts from KS 1731-260\label{burst}}
\tablehead{
\colhead{Burst} & \colhead{Date} & \colhead{$t_{rise}$} 
& \colhead{$t_s$} & \colhead{$t_d$} & 
\colhead{$F_{peak}$} & \colhead{$E_{b}$}  
&\colhead{$\tau$} & \colhead{Radius} \\
\colhead{Number} & \colhead{UT} & \colhead{(s)} 
& \colhead{(s)} & \colhead{(s)} 
& \colhead{($10^{-8}$ \ergcms)\tablenotemark{a}} & 
\colhead{($10^{-7}$ erg cm$^{-2}$)} & \colhead{(s)} & \colhead{Expansion}
}
\startdata
1 & 1996 Jul 14 04:23:42 & 3.25 & 4.0 & 4.08(8) & 4.6(1) &  & & \\
  &                   &      & 11.0 & 9.4(6) &      & 3.24(7) & 6.6(2) & Y \\ 
2 & 1997 Oct 29 08:10:23 & 3.00 & 4.5 & 4.4(1)  & 6.3(1) &   & & \\
  &		      &      & 10.0 & 11.4(3)&     & 4.25(7) & 6.8(2) & Y\\
3 & 1998 Oct 02 07:34:54 & 3.75 & 9.0 & 11.0(1) & 4.4(1) & 7.01(7) & 15.9(4) &N\\
4 & 1998 Oct 02 14:00:27 & 3.50 & 9.0 & 7.6(2) & 4.7(1) &  & & \\
  &                   &      & 20.0 & 9.4(2) &    & 7.02(8) & 14.6(4) & N \\
5 & 1998 Oct 03 09:11:55 & 5.25 & 9.0 & 11.5(1) &3.4(1) & 5.85(7) & 17.2(5) &N \\
6 & 1998 Oct 06 03:26:58 & 2.75 & 9.0 & 9.11(6) &4.4(1) & 6.78(7) & 15.1(4)& N \\
7 & 1999 Feb 23 03:06:16 & 1.25 & 1.5 & 3.7(1) & 2.1(1) & 0.77(2) & 3.3(2) & N \\
8 & 1999 Feb 26 17:10:53 & 1.50 & 5.0 & 4.01(9) &5.3(1) &  & & \\
  &                   &      & 10.0 & 9.2(2) &     & 4.57(7) & 8.6(2) & Y \\
9 & 1999 Feb 27 17:23:01 & 2.00 &  2.5 & 6.27(7) &5.2(1) &       &  & \\
  &                   &      & 15.0 & 14.6(9) &    & 4.65(9) & 8.6(2) & Y \\
\enddata
\tablenotetext{a}{Bolometric flux (see text).}
\tablecomments{The three times listed are the rise time of the burst 
($t_{rise}$), the approximate starting time of the exponential decay 
measured from the start of the burst ($t_s$), and the $e$-folding time for
the exponential decay of the burst ($t_d$).}
\end{deluxetable}

\begin{deluxetable}{lcccccc}
\scriptsize
\tablecolumns{7}
\tablewidth{0pc}
\tablecaption{Oscillations in Bursts from \ksxrb\label{osc}}
\tablehead{
\colhead{Burst} & \colhead{Fundamental} & 
\colhead{Harmonic} & 
\colhead{Sub-Harmonic} & \colhead{Time\tablenotemark{a}} & 
\colhead{Signal} 
& \colhead{Radius} \\
\colhead{Burst} & \colhead{(\% RMS)} & 
\colhead{(\% RMS)} & 
\colhead{(\% RMS)} & \colhead{Signal (s)} & \colhead{Duration (s)} 
& \colhead{Expansion} 
}
\startdata
1 & $9.2 \pm 0.6$  & $<$11.1 & $<$5.9 & 2 & 3 & Y \\
2 & $5.4 \pm 1.1$ & $<$6.9 & $<$6.1 & 4 & 2 & Y \\
3 & $<$9.2 & $<$9.2 & $<$9.9 & \nodata & \nodata & N \\
4 & $<$6.6 & $<$8.0 & $<$8.7 & \nodata & \nodata & N \\
5 & $<$8.9 & $<$7.8 & $<$9.0 & \nodata & \nodata & N \\
6 & $<$8.0 & $<$7.2 & $<$6.7 & \nodata & \nodata & N \\
7 & $8.6 \pm 1.1$ & $<$13.3 & $<$8.9 & 0 & 1 & N \\
8 & $6.5 \pm 1.5$ & $<$5.6 & $<$11.5 & 3 & 1 & Y \\
9 & $12.6 \pm 0.4$ & $<$5.6 & $<$7.8 & 2 & 6 & Y \\
\enddata
\tablenotetext{a}{Measured from the start time of the burst; see Table~2.}
\tablecomments{Values with one-sigma error estimate represent percent 
    RMS of the largest detected signal; upper limits represent the 90\% 
    confidence level of the percent RMS for non-detections.}
\end{deluxetable}

\begin{deluxetable}{ccrrrrrc}
\scriptsize
\tablecolumns{8}
\tablewidth{0pc}
\tablecaption{Frequency Evolution in Bursts from \ksxrb\label{model}}
\tablehead{
\colhead{Burst} & \colhead{$\Delta t$ (s)} & \colhead{\phiz\ (cyc.)} & 
\colhead{$\nu_0$ (Hz)} & \colhead{$\dot{\nu}_0$ (Hz s$^{-1}$)} & 
\colhead{$\ddot{\nu}_0$ (Hz s$^{-2}$)} & 
\colhead{\dddnuz\ (Hz s$^{-3}$)} & 
\colhead{$\chi^2$/dof} }
\startdata
 1  & ~2.00--11.75 & 0.41$\pm$0.05  &  523.67$\pm$0.06
    &  0.20$\pm$0.04  &  $-$0.013$\pm$0.004  &  ~
    &  9.5/8   \\
 2  & ~2.5--10.5   & 0.36$\pm$0.05  &  523.90$\pm$0.14
    &  $-$0.30$\pm$0.16  &  0.15$\pm$0.04  &  $-$0.011$\pm$0.004  
    &  5.0/5   \\
 7  & 0.25--1.00 &  0.52$\pm$0.08  &  523.69$\pm$0.20 & $<$4.5 & ~ & ~
    &  7e-5/1  \\
 8  &  0.00--0.75  &  0.98$\pm$0.07  &  522.63$\pm$0.21 & $<$5.4 & ~ & ~
    &  0.01/1  \\
 8  &  2.0--5.0  &  0.88$\pm$0.07  &  524.76$\pm$0.13 
    &  $-$0.72$\pm$0.14  &  $<$4.6  &  ~ 
    &  7.0/2   \\
 9  & ~2.0--11.0 &  0.79$\pm$0.05  &  523.54$\pm$0.04  
    &  0.25$\pm$0.03  &  $-$0.019$\pm$0.003  &  ~
    &  13.3/11 \\
\enddata
\tablecomments{Polynomial fits to the phase models for bursts with
    detected burst pulsations.  Time $\Delta t$ is measured from the
    burst start (Table~2), uncertainties are one-sigma, and
    quoted upper limits are two-sigma limits on the absolute value of
    the coefficient.}
\end{deluxetable}

\begin{deluxetable}{lrrrrrc}
\scriptsize
\tablecolumns{7}
\tablewidth{0pc}
\tablecaption{Frequency Evolution: Exponential Models\label{expmodel}}
\tablehead{
\colhead{Burst}  &  \colhead{\phiz\ (cyc.)}  &  \colhead{\nua\ (Hz)} 
    & \colhead{\taunu\ (s)}  &  \colhead{$\Delta\nu$ (Hz)}
    & \colhead{$\Delta\phi$ (cyc.)}  &  \colhead{$\chi^2$/dof} }
\startdata
 1  &  0.41$\pm$0.04        &  524.61$^{+0.13}_{-0.07}$  
    &  4.1$^{+1.3}_{-0.9}$  &  0.96$\pm$0.04  
    &  3.9$^{+3.4}_{-1.1}$  &  10.8/8  \smstrut \\  
 9  &  0.84$\pm$0.06        &  524.48$^{+0.05}_{-0.03}$ 
    &  2.6$^{+0.4}_{-0.3}$  &  1.06$\pm$0.05 
    &  2.8$^{+0.7}_{-0.4}$  &  13.2/11 \smstrut \\
\enddata
\tablecomments{Exponential-relaxation fits to the phase models for
    Bursts 1 and 9.  Time intervals are the same as in
    Table~4.  $\Delta\phi$ is the total phase loss relative
    to the asymptotic frequency, and is not an independent parameter of
    the fit (see text for details).}
\end{deluxetable}

\end{document}